\documentclass[11pt]{article}
\usepackage{asherspector}

\usepackage[margin=0.75in]{geometry}
\usepackage{color}
\usepackage{graphicx}
\usepackage{subfig}
\usepackage[font=small,skip=0pt]{caption}

\usepackage{rotating}
\usepackage{verbatim}
\usepackage{hyperref}
\usepackage[normalem]{ulem}
\usepackage[authoryear]{natbib}
\usepackage{bbm}
\usepackage{array}
\usepackage{float}
\usepackage{algorithm}
\usepackage[noend]{algpseudocode}

\usepackage[flushleft]{threeparttable}
\usepackage{tablefootnote}

\usepackage[perpage]{footmisc}

\newcommand{\ba}{\bs{a}}

\newcommand{\slope}{\widetilde{\mathrm{\rho}}}

\newcommand{\mosaic}{_{\mathrm{mosaic}}}
\numberwithin{equation}{section}

\newcommand{\tildemuKm}{\tilde\mu_K^{(m)}}
\newcommand{\KS}{\mathrm{KS}}
\newcommand{\singleton}{\mcS}
\newcommand{\nonsingleton}{\singleton^C}
\newcommand{\degree}{\texttt{degree}}
\newcommand{\support}{\texttt{support}}

\newcommand{\bigsupport}{\texttt{S}}
\newcommand{\bigdegree}{\texttt{D}}
\newcommand{\upM}{^{(M)}}
\newcommand{\optM}{^{(\star,M)}}
\newcommand{\upm}{^{(m)}}
\newcommand{\sgconstant}{C_{\mathrm{subG}}}
\newcommand{\reverse}{^{\mathrm{reverse}}}

\newcommand{\pval}{p_{\mathrm{val}}}
\newcommand{\aug}{^{\mathrm{trans}}}
\newcommand{\MosaicResid}{\texttt{MosaicResid}}
\newcommand{\MosaicRandomize}{\texttt{MosaicRandomize}}
\newcommand{\hatse}{\hat\sigma}
\newcommand{\theor}{^\mathrm{T}}
\newcommand{\emp}{^\mathrm{E}}

\newtheorem{assumptionalt}{Assumption}[assumption]
\newenvironment{assumptionp}[1]{
  \renewcommand\theassumptionalt{#1}
  \assumptionalt
}{\endassumptionalt}

\title{Mosaic inference on panel data}

\author{Asher Spector \thanks{Department of Statistics, Stanford University} \and Rina Foygel Barber \thanks{Department of Statistics, University of Chicago} \and Emmanuel  Cand{è}s \footnotemark[1] \thanks{Department of Mathematics, Stanford University}}
\date{\today\thanks{The authors thank John Cherian, Kevin Guo, Lihua Lei, Ginnie Ma, Yash Nair, and Joe Romano for their valuable suggestions. A.S. was partially supported by the Two Sigma Graduate Fellowship Fund,
the Citadel GQS PhD Fellowship, and a Graduate Research Fellowship from the National Science Foundation. R.F.B. was partially supported by the National Science Foundation via grant DMS-2023109, and by the Office of Naval Research via grant N00014-24-1-2544. E.J.C. was supported by the Office of Naval Research grant N00014-20-1-2157, the National Science Foundation grant DMS-2032014, and the Simons
Foundation under award 814641.}}

\begin{document}
\maketitle

\begin{abstract}
Analysis of panel data via linear regression is widespread across disciplines. To perform statistical inference, such analyses typically assume that clusters of observations are jointly independent. For example, one might assume that observations in New York are independent of observations in New Jersey. Are such assumptions plausible? Might there be hidden dependencies between nearby clusters? This paper introduces a mosaic permutation test that can (i) test the cluster-independence assumption and (ii) produce confidence intervals for linear models without assuming the full cluster-independence assumption. The key idea behind our method is to apply a permutation test to carefully constructed residual estimates that obey the same invariances as the true errors. As a result, our method yields finite-sample valid inferences under a mild ``local exchangeability" condition. This condition differs from the typical cluster-independence assumption, as neither assumption implies the other. Furthermore, our method is asymptotically valid under cluster-independence (with no exchangeability assumptions). 
Together, these results show our method is valid under assumptions that are arguably weaker than the assumptions underlying many classical methods. In experiments on well-studied datasets from the literature, we find that many existing methods produce variance estimates that are up to five times too small, whereas mosaic methods produce reliable results. We implement our methods in the python package \texttt{mosaicperm}.





\end{abstract}


\section{Introduction}\label{sec::intro}


\subsection{Setting}\label{subsec::contribution}


Panel datasets are ubiquitous across the social and natural sciences. While not without its own limitations \citep{arkhangelsky_imbens_causalsurvey2024}, linear regression via ordinary least squares (OLS) remains among the most popular tools for analysis of panel data, including common methods such as two-way fixed effects models and difference-in-difference analyses as special cases \citep{mackinnon2023clusterrobust}.

This paper develops methods to test and weaken the conventional assumptions needed to perform inference on linear regression models of panel data. For units $i=1, \dots, N$ at times $t=1, \dots, T$, we assume we observe outcomes $Y_{i,t}$ obeying the linear model below:
\begin{equation}\label{eq::linear_model}
    Y_{i,t} = X_{i,t}\trans \beta\opt + \epsilon_{i,t},
\end{equation}
where $X_{i,t} \in \R^D$ are covariates, $\beta\opt \in \R^D$ are nonrandom coefficients, and $\epsilon_{i,t} \in \R$ are random errors. Throughout, we treat $X_{i,t}$ as fixed (nonrandom) controls; note that $X_{i,t}$ may contain dummy variables to incorporate fixed effects. As notation, $\epsilon \in \R^{N \times T}$ denotes the matrix of errors, and for any $G \subset [N]$, $\epsilon_{G} \in \R^{|G| \times T}$ denotes the submatrix of errors from the units in $G$. 

Statistical inference on (e.g.) the first coefficient $\beta_1\opt$ is challenging because the observations may not be independent---indeed, panel data often exhibit cross-sectional dependence among units as well as autocorrelation across time. A typical solution is to assume that for a known ``clustering" or partition $C_1, \dots, C_M \subset [N]$ of units, the residuals $\epsilon_{C_1}, \dots, \epsilon_{C_M}$ are jointly independent.\footnote{Other ways of clustering observations are possible, e.g., clustering by the observation time. Our methods and theory extend to other partitions of observations, but we focus on clustering across units since this case is common in empirical practice. Multi-way clusterings (as discussed in, e.g.,  \cite{cameron2014multiway}) are beyond the scope of this paper.} This assumption facilitates statistical inference whenever the errors are mean-zero and the number of observations or clusters is sufficiently large; Assumption \ref{assump::conventional} (S for ``standard") states these conditions below.

\begin{assumptionp}{S}\label{assump::conventional} The errors are mean-zero and cluster independent, i.e., $\E[\epsilon] = 0$ and $\{\epsilon_{C_m}\}_{m=1}^M$ are jointly independent. Also, the number of clusters $M$ is large, enabling asymptotic inference (see below).
\end{assumptionp}

\begin{remark} The precise assumptions in Assumption \ref{assump::conventional} can be relaxed. For example, many existing results only require that $\{\epsilon_{C_m}\}_{m=1}^M$ are jointly uncorrelated. Others do not require mean-zero residuals and only assume that (e.g.) $\E[\sum_{i,t} \epsilon_{i,t} X_{i,t}] = 0$. We state Assumption \ref{assump::conventional} mainly to give a conceptual overview without focusing on technical details.
\end{remark}

A large literature has developed methods for inference under assumptions such as Assumption \ref{assump::conventional} \citep[e.g.,][]{white_1984, LZ_1986, Hansen_2007, Djogbenou_2019, HansenLee_2019}. Other works show how to perform inference when $M$ is small but the number of observations per cluster is large \citep[e.g.,][]{BCH_2011, IM_2010, ibragimov2016}. A common theme is that asymptotic results can depend on the aspect ratio $N / T$ \citep{Hansen_2007}, the number of clusters, and the (unobservable) degree of heterogeneity within and between clusters \citep{Djogbenou_2019, HansenLee_2019, mackinnon2023clusterrobust}. Furthermore, asymptotics often fail in common settings such as when one cluster is overly large \citep{HS_2020} or when a regressor of interest has nonzero variation within only a few clusters \citep{MW-EJ-2018}. Another line of literature shows that bootstrap-based methods can partially (but not fully) solve these problems \cite[e.g.,][]{CGM_2008, CM_2015, Djogbenou_2019, MNW-bootknife-2022, mackinnon2023clusterrobust}. 

\subsection{Contribution}

While the literature above has developed methods to perform inference under Assumption \ref{assump::conventional} (and other asymptotic regimes), the goal of our paper is to introduce methods to test and weaken this assumption. Throughout, we focus on obtaining finite-sample validity results in addition to asymptotic ones, since asymptotic results for panel data can depend on the choice of asymptotic regime.

\textbf{Contribution 1: testing cluster-independence.} We introduce a \textit{mosaic permutation test} to test cluster independence:
\begin{equation}\label{eq::null}
    \mcH_0 : \epsilon_{C_1}, \dots, \epsilon_{C_M} \text{ are jointly independent.} 
\end{equation}
Our test has three key properties. First, it is exactly valid in finite samples under a mild local exchangeability condition on the errors. Second, it permits the use of a wide set of test statistics---including those based on machine learning techniques---to detect violations of the cluster-independence assumption. Third, when using a natural quadratic test statistic, our test asymptotically controls the Type I error rate under Assumption \ref{assump::conventional}, even when the local exchangeability condition completely fails. In short, our test is exactly valid under a local invariance condition and asymptotically valid under ``conventional" assumptions. 

\textbf{Contribution 2: inference.} We show how to invert our test to produce confidence intervals for linear models. These intervals have four key properties. First, they have a transparent formula reminiscent of a block-bootstrap-like procedure. Second, they can be computed efficiently---typically at roughly the computational cost of computing ordinary least squares (OLS) residuals. Third, they are valid in finite samples under a joint local exchangeability condition. Joint local exchangeability and cluster-independence are non-nested assumptions (neither implies the other), but empirically, we find that methods based on joint local exchangeability yield more reliable inferences. Fourth, our confidence intervals are still valid asymptotically under ``standard" assumptions (Assumption \ref{assump::conventional}) even when the local exchangeability condition fails. Thus, our method is valid under assumptions that are strictly weaker than the ``standard" assumptions (Assumption \ref{assump::conventional}).

\textbf{Validation through empirical diagnostics.} All assumptions in this paper---cluster independence, local exchangeability, mean-zero errors, and more---are only approximations of reality and will not hold exactly in practice. Thus, we perform empirical diagnostics on real data to assess the reliability of inferential procedures. 

These diagnostics, based on data splitting, are described in detail in Section \ref{sec::experiments}, but we present the main idea here. Given a dataset, we randomly split it into two disjoint folds. On each fold, we compute coefficient estimates, standard errors, and confidence intervals. We then run two diagnostics. First, if the standard errors are accurate, then the squared sum of the standard errors should be an asymptotically unbiased estimate of the squared distance between the two estimators \citep[e.g.,][]{ibragimov2016}. Second, if existing asymptotic theories are correct, we can explicitly compute the probability that two confidence intervals (computed on disjoint folds of the data for the same parameter) overlap. We then compare this theoretical probability to the empirical probability that the confidence intervals overlap. 

\begin{figure}
    \centering
    \includegraphics[width=0.8\linewidth]{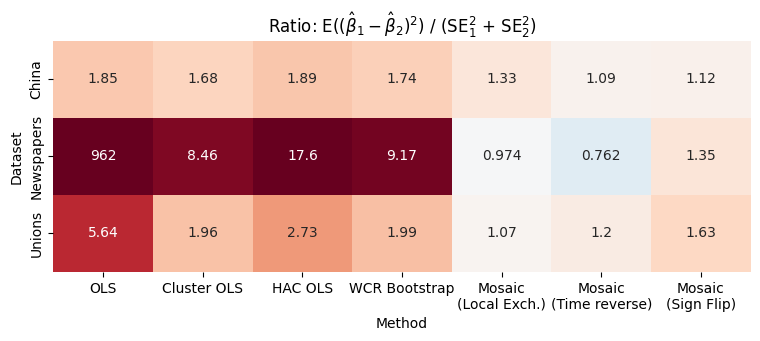}
    \caption{For each dataset, we split the data into two folds and compute two estimators and two standard errors for the same coefficient using various statistical methods. For each method and dataset, this figure shows the average ratio of the squared distance between the two estimators divided by the squared sum of the standard errors. (Results are averaged over (a) many random splits of the same dataset and (b) several choices of the feature of interest, since the papers in question analyze regression coefficients for multiple features.) Roughly speaking, values below one (shown in blue) yield conservative inference, whereas values above one (shown in red) yield anticonservative inference.}\label{fig::ratios}
\end{figure}

\begin{figure}[!h]
    \centering
    \includegraphics[width=\linewidth]{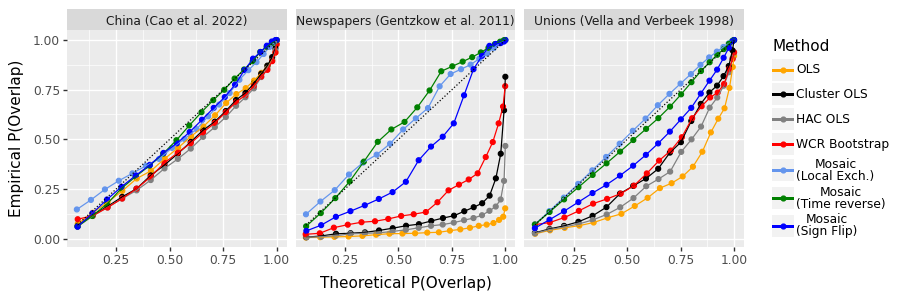}
    \caption{For each dataset, we split the data into two folds and compute two level $\alpha$ confidence intervals for the same coefficient (using various statistical methods). Under classical asymptotics, these confidence intervals should overlap with an easily computable ``theoretical probability." This figure plots the theoretical probability of overlap $\hat p\theor(\alpha)$ and the empirical probability of overlap $\hat p\emp(\alpha)$. The results are averaged over (a) many random splits of the same dataset and (b) several different choices of the feature of interest. It shows that these theoretical and empirical probabilities do not line up well for existing methods. For example, when $\alpha=0.05$, theory predicts that these confidence intervals should overlap with probability $\approx 99.5\%$, but WCR bootstrap intervals only overlap $60\%$ of the time for the \cite{gentzkow2011newspaper} dataset (which is what we would expect to see if we set $\alpha \approx 0.71$). In contrast, mosaic confidence intervals (especially when based on local exchangeability) have improved performance.}\label{fig::coverage_results}
\end{figure}

We analyze three panel datasets from the economics literature and perform these two empirical tests. Figure \ref{fig::ratios} shows that classical and state-of-the-art methods based on the cluster independence assumption produce standard errors that are much too small, and Figure \ref{fig::coverage_results} shows that the confidence intervals from these methods likely severely undercover. In contrast, mosaic intervals produce more reliable results. These results give some reassurance that the local exchangeability assumption yields reasonable inferences in practical settings.

\subsection{Practical implications of our empirical results}

This section gives two distinct interpretations of our main empirical results. Under either interpretation, our method's strong performance highlights its practical value. After discussing this topic, we connect our work to a broader discussion on model-based and design-based inference.

Our empirical diagnostics ask:
\begin{quote}
    \emph{When we apply a statistical method to two subsets of the same dataset, are the results compatible with one another? For example, do the confidence intervals from these subsets overlap?
    }
\end{quote}
Our results show that many existing methods yield incompatible results, whereas our mosaic methods yield greater empirical consistency. We argue that empirical consistency is valuable and incompatible results are practically dissatisfying for at least one of two reasons.

\emph{Reason 1: inaccurate assumptions or unreliable methodology.} The first possible explanation is that our methodology or assumptions are wrong. For example, perhaps a method falsely assumes the errors are cluster-independent, making the standard errors too small. Alternatively, perhaps a method relies on asymptotic approximations that are not accurate in this setting.

\emph{Reason 2: model misspecification.} An alternative explanation is that the estimand differs across folds. For example, consider a \emph{univariate} regression of log-income on education in China (with no fixed effects). A confidence interval based on data from all of China might differ substantially from intervals calculated separately for Northern and Southern regions, perhaps because the estimand differs in Northern and Southern China. However, this suggests that the initial model is misspecified, since the original univariate regression $Y_{i,t} = X_{i,t} \beta\opt + \epsilon_{i,t}$ does not explicitly allow for heterogeneous effects across geography. Indeed, it is well-known that this type of model misspecification in regressions can lead to misleading results: without accounting for heterogeneous effects, a final effect estimate is not necessarily an interpretable average of region or individual-level effects \citep{twfe2020aer, chen2025potentialweightsimplicitcausal}.


In either scenario, we argue that reporting results from methods that yield incompatible outcomes across subsets of the same dataset is unsatisfying. At best, such results mask meaningful heterogeneity within the data that the model does not account for; and at worst, they are genuinely misleading. This highlights the practical value of our proposed confidence intervals, which are empirically more likely to produce compatible results. Furthermore, all methods in our experiments use the same linear model, meaning that model misspecification (reason 2) does not explain performance gaps between methods. Rather, this suggests that, in our experiments, mosaic methods better quantify uncertainty than existing methods due to differing inferential assumptions. 

\begin{remark}[Design-based inference methods need not produce compatible results.] Our previous arguments only apply to ``model-based" approaches that make assumptions about the joint law of the errors. In contrast, \textit{design-based} procedures assume either that (i) clusters are randomly sampled from a finite population or (ii) a treatment is randomly assigned \citep{abadie_2020_regression, abadie2022clustering, rambachan_roth_2020_designbased, arkhangelsky_imbens_causalsurvey2024}. Unlike model-based procedures, design-based procedures intentionally produce ``incompatible" results when run on different subsets of the same dataset, since the estimand depends transparently on a finite population of individuals. Thus, we strongly recommend design-based procedures whenever assumptions (i) or (ii) hold. 

However, in many empirical applications, researchers simply use whatever data is available, and the regressor of interest has not obviously been randomized.  Our paper focuses exclusively on these settings, where researchers typically use \textit{model-based} approaches that make assumptions about the joint law of the errors. These mathematical assumptions---for example cluster independence---are (provably) testable. 
\end{remark}

\subsection{Additional related literature}

Our method builds on several lines of literature. First, it builds on recent work on finite-sample valid permutation tests for regression \citep{lei2020cpt, wen2022ResidualPT, dhaultfoeuille2024ARP, guan2024palmrt, pouliot2025}. However, a main conceptual difference is that we aim to establish finite-sample results under only mild restrictions on the joint law of the errors. In contrast, \cite{lei2020cpt, wen2022ResidualPT, guan2024palmrt} provide finite-sample guarantees assuming that all errors are jointly exchangeable or i.i.d. \cite{dhaultfoeuille2024ARP} require the joint support of all controls to be small relative to the number of observations. Furthermore, \cite{pouliot2025} requires an invariance based on block permutations which is assumed to increase power and facilitate analytical analysis but not necessarily to increase robustness to dependence between the observations. While \cite{dhaultfoeuille2024ARP, pouliot2025} prove their tests are asymptotically robust to heteroskedasticity, these results assume the observations are independent, which is unrealistic for panel data. We also note that our method allows the use of a broader range of test statistics than aforementioned works. 
Lastly, we build on the mosaic permutation test for factor models from \cite{mosaic_factor_paper2024}, adapting it to the panel data setting and introducing a new set of invariance assumptions that ensure asymptotic robustness—--something the original test lacks. There are several other key methodological differences, but for now, we defer these to Remark \ref{rem::augment_model}.

Second, our asymptotic results build on literature showing the robustness of permutation tests \citep[e.g.,][]{romano1990, neuhaus1993, janssen_pauls_2003, janssen_thorsten_2005, chung_romano_2013, chung_romano_2016_multivariate, chung_romano_2016_ustats, dhaultfoeuille2024ARP, pouliot2025}. A key challenge, distinct from previous work, is that the asymptotic distribution of our test statistic (a degenerate U-statistic) is non-universal; it depends on the aspect ratio $\frac{N}{T}$ and the error distribution \citep{bhattacharya2022asymptoticdistributionrandomquadratic}. Using a novel method-of-moments technique, we show that our test automatically recovers the limiting distribution and provides asymptotic Type I error control under standard assumptions (Assumption \ref{assump::anticoncentration}), even when our core invariance assumptions are false. Our approach has several advantages: it avoids studentizing or prepivoting the statistic \citep[e.g.,][]{chung_romano_2013}, making it suitable for panel data where consistent variance estimation is challenging; and, unlike recent analyses of degenerate U-statistics \citep[e.g.,][]{menzel2021}, it does not require (but nor does it prohibit) the number of time-points $T$ to diverge asymptotically. Lastly, unlike \cite{CRS_2017, canay_permutation_tests2017}, our results do not require the invariance assumption to hold even asymptotically. That said, our approach only applies to a specific class of quadratic test statistics. Also, a weakness relative to \cite{canay_permutation_tests2017, CRS_2017} is that our asymptotic results require the number of clusters to diverge. Instead, when the number of clusters is small, we rely on finite-sample validity results.

Third, three recent works \citep{ibragimov2016, Cai_2021, MNW-testing-2023} also test the cluster independence assumption. Indeed, one of our empirical validation schemes is related to the test statistic from \cite{ibragimov2016}. However, each of these works is designed to test the null of a fine clustering against the alternative of a coarser clustering, i.e., testing whether clustering at the county or state level is more appropriate. Furthermore, each of these works relies on asymptotic arguments that assume either that the number of fine clusters is large \citep{ibragimov2016, MNW-testing-2023} or that a central limit theorem applies within each fine cluster \citep{Cai_2021}. In contrast, we focus on testing whether the cluster independence assumption holds for a potentially coarse clustering, and our main validity results hold in finite samples. 

Fourth, our work contributes to a literature that uses randomization-based methods to perform inference on panel data. However, the majority of these works assume that cluster-level estimates of the parameters of interest are asymptotically Gaussian or symmetric \citep{CRS_2017, CCKS_2021, CSS_2021, Hag_2019b}, yielding asymptotic validity. However, it can be challenging to verify whether within-cluster asymptotics are accurate, especially when there is not much variation among a covariate of interest. 
For this reason, we introduce a test that is finite-sample valid under an interpretable (but non-trivial) local exchangeability assumption about the errors.

\section{Local exchangeability and other invariance assumptions}\label{sec::invariances}

This section introduces the core invariance assumption used in our paper. To maximize the transparency of our finite-sample results, we make assumptions directly about the errors rather than assuming that (e.g.) a test statistic is approximately Gaussian or symmetric. We state our main assumption below.

\begin{assumptionp}{MI}[Marginal Invariance]\label{assump::marginv} Given a partition or ``clustering" $C_1, \dots, C_M \subset [N]$ of units, there exists a known transformation matrix $P \in \R^{T \times T}$ such that for all $m \in [M]$,
\begin{equation}
    \epsilon_{C_m} \disteq \epsilon_{C_m} P.
\end{equation}
Furthermore, $P$ is (1) symmetric, so $P\trans = P$, and (2) idempotent, so $P^2 = I_T$. 
\end{assumptionp}

If the clusters of errors $\{\epsilon_{C_m}\}_{m=1}^M$ are also jointly independent, Assumption \ref{assump::marginv} implies the joint invariance assumption below. 
\begin{assumptionp}{JI}[Joint Invariance]\label{assump::jointinv} For any $z_1, \dots, z_M \in \{0,1\}$,
\begin{equation*}
    (\epsilon_{C_1}, \dots, \epsilon_{C_M}) \disteq (\epsilon_{C_1} P^{z_1}, \dots, \epsilon_{C_M} P^{z_M}).
\end{equation*}
In words, we may separately choose whether to transform (or not) each cluster of errors without changing the joint law of the errors.  
\end{assumptionp}

If both Assumptions \ref{assump::marginv} and the cluster-independence assumption hold, they imply 
Assumption \ref{assump::jointinv}. However, in and of itself, Assumption \ref{assump::jointinv} is neither weaker nor stronger than the standard cluster independence assumptions.

We now give three examples of this invariance assumption. Our default choice is local exchangeability (Invariance \ref{invariance::local_exch}), but all three can be reasonable in different contexts. We acknowledge that these assumptions may not be perfectly accurate---no assumptions are---but (1) we think that they are substantively plausible, (2) we find that methods based on these assumptions outperform competitors in Section \ref{sec::experiments}, and (3) we will soon prove that our methods are asymptotically robust to near-arbitrary violations of these assumptions under classical assumptions. 

\begin{figure}[!h]
    \includegraphics[width=\linewidth]{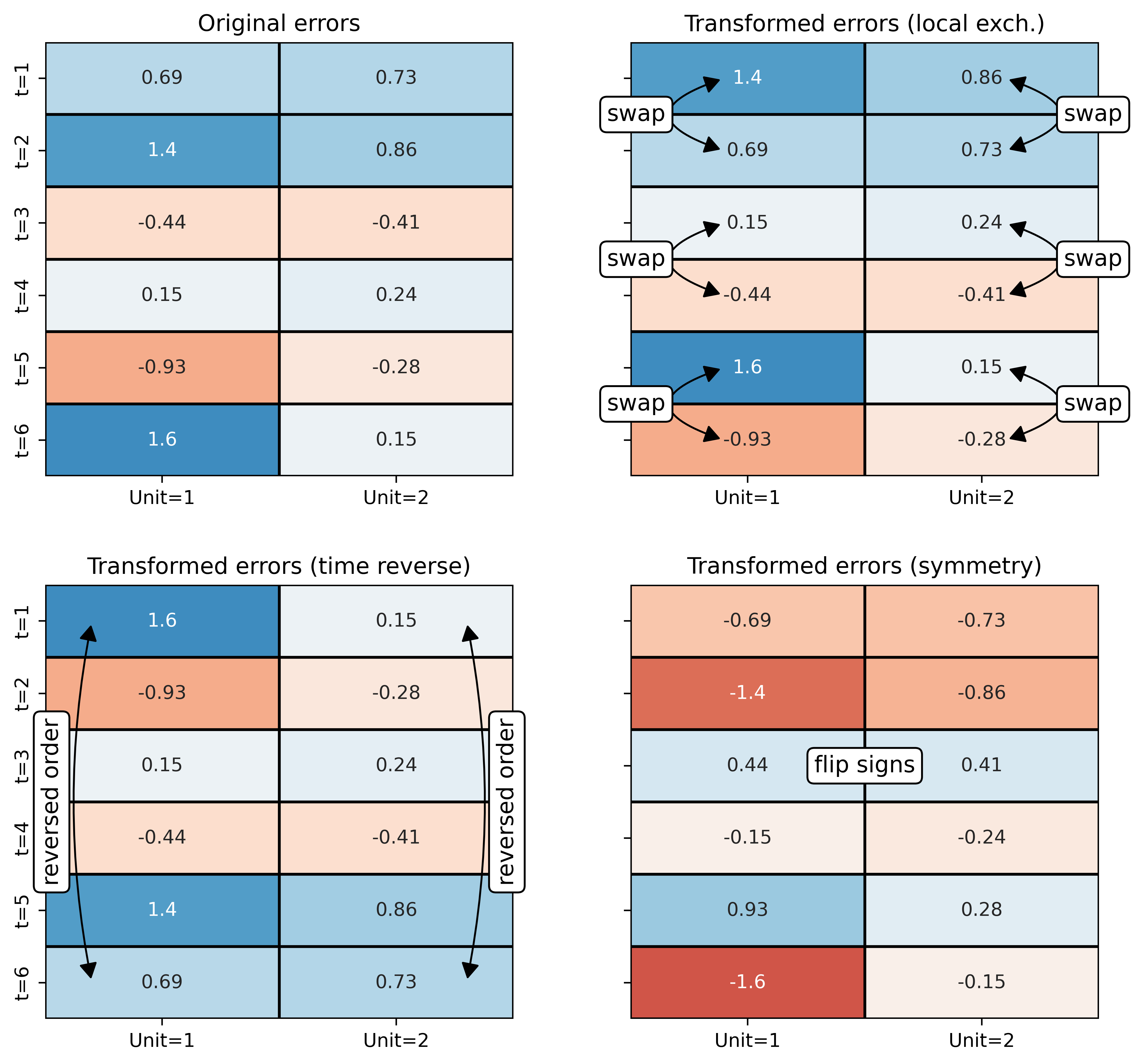}
    \caption{This figure illustrates three examples of the marginal invariance assumption with $|C_m| = 2$ units and $T=6$ time periods. To ease readability, cells are colored by their values.
    }\label{fig::invariance_combined}
\end{figure}

\begin{invariance}[Symmetry] Symmetry takes $P = - I_T$. It asserts that each cluster of errors is symmetric:
\begin{equation}
    \epsilon_{C_m} \disteq - \epsilon_{C_m}.
\end{equation}
\end{invariance}
This invariance is simple and easy to interpret. However, it may not always be the most plausible since the errors may not be exactly symmetric (or even exactly mean zero).

\begin{invariance}[Time reversal] Time reversal asserts that $\epsilon_{C_m}$ is invariant to reversing the temporal order of the observations, as illustrated in Figure \ref{fig::invariance_combined}.
Formally, let $P\reverse \in \{0,1\}^{T \times T}$ be the permutation matrix encoding the time-reversal permutation, so $P\reverse_{t_0,t_1} = 1$ if $t_0 + t_1 = T$ and $P\reverse_{t_0,t_1} = 0$ otherwise. Then
\begin{equation}
    \epsilon_{C_m} \disteq \epsilon_{C_m} P\reverse.
\end{equation}
\end{invariance}

\begin{invariance}[Local exchangeability]\label{invariance::local_exch} For each odd time $t < T$, we say that the adjacent times $t$ and $t+1$ are \textit{paired}. (If $T$ is odd, the last observation $T$ is paired with itself.) Let $P\swap \in \{0,1\}^{T \times T}$ be the permutation matrix which swaps paired times---i.e., $P_{k,\ell}\swap = 1$ if $(k, \ell)$ are paired.
Then, we assume
\begin{equation}
    \epsilon_{C_m} \disteq \epsilon_{C_m} P\swap.
\end{equation}
\end{invariance}
In words, for each unit, $\epsilon_{C_m} P\swap$ swaps the errors in each adjacent pair of observations. This idea is illustrated in Figure \ref{fig::invariance_combined}.

\begin{figure}[!h]
\begin{centering}
\includegraphics[width=0.6\linewidth]{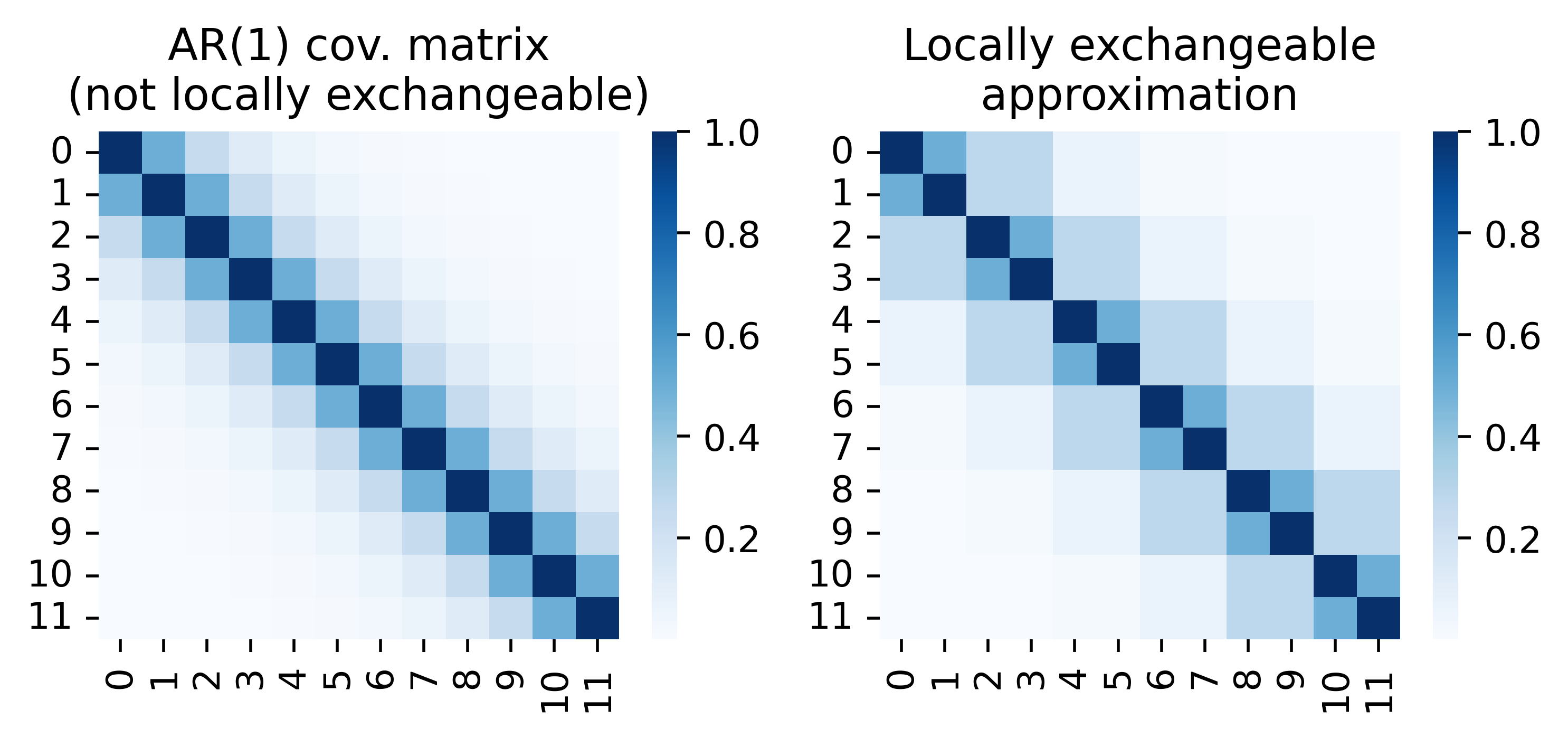}
\caption{The left plot shows an AR(1) covariance matrix with $T=12$ timepoints. The right plot shows a locally exchangeable approximation to this covariance matrix.}\label{fig::ar1}
\end{centering}
\end{figure}

Below, we discuss the restrictions implied by local exchangeability. 
\begin{itemize}[itemsep=0.5pt, topsep=0pt, leftmargin=*]
\item \textit{Cross-sectional dependence}: local exchangeability allows arbitrary cross-sectional dependence across units within clusters, since all observations at time $t$ are swapped jointly with observations at time $t+1$.
\item \textit{Heterogeneity across units}: errors are only swapped within units, so local exchangeability allows the law of $\epsilon_i \in \R^T$ (the time-series of errors for unit $i$) to differ arbitrarily across units. 
\item \textit{Local stationarity}: local exchangeability requires the errors to be locally stationary, i.e., it implies $\epsilon_{i,t} \disteq \epsilon_{i,t+1}$ for odd $t$. This will hold approximately if the law of $\epsilon_{i,t}$ changes slowly over time. Furthermore, even if there are a few sharp changes in the law of $\epsilon_{i,t}$ (as a function of $t$), these changes will each only affect one ``swap," since local exchangeability permits the law of $\epsilon_{i,t+2}$ to differ arbitrarily from the law of $\epsilon_{i,t}$.
\item \textit{Autocorrelation}: local exchangeability mildly restricts the autocorrelation pattern of the errors. E.g., suppose $\epsilon_{i,t} = \rho \epsilon_{i,t-1} + \sqrt{1-\rho^2} \zeta_{i,t}$ is an AR(1) process based on i.i.d.~random variables $\{\zeta_{i,t}\}_{t=1}^T$.
This violates local exchangeability, because then (e.g.) $(\epsilon_{i,1}, \epsilon_{i,2}, \epsilon_{i,3}) \not \disteq (\epsilon_{i,2}, \epsilon_{i,1}, \epsilon_{i,3})$. However, AR(1) processes can be imperfectly approximated by locally exchangeable processes (see Figure \ref{fig::ar1}).
\end{itemize}

We note also that the \textit{joint} local exchangeability (Assumption \ref{assump::jointinv}) is simply different---neither weaker nor stronger---than a traditional cluster-independence assumption. Indeed, it is easy to think of examples where one assumption holds but not the other. For example, imagine that $\epsilon_{i,t} \simind \mcN(0,t)$: here, cluster independence certainly holds, but local exchangeability fails since the variance of each residual depends on $t$ (although it arguably holds approximately since the variance is a smooth function of $t$). Alternatively, imagine that unbeknownst to the researcher, a single disturbance contaminates the errors at the first $T_0$ timepoints:
\begin{equation*}
    \epsilon_{i,t} = \gamma_{i,t} + \I(t \le T_0) \lambda,
\end{equation*}
where (e.g.) $\gamma_{i,t}, \lambda \simiid \mcN(0,1)$. In this case, cluster independence fails dramatically, but local exchangeability holds exactly when $T_0$ is even, and it holds approximately when $T_0$ is odd (since for each unit, only one ``swap" is not exchangeable).

Overall, this intuition suggests that local exchangeability may hold (at least approximately) for many real-world datasets. 
Our empirical results (Section \ref{sec::experiments}) support this intuition.

\section{Testing cluster independence}\label{sec::mpt}

\subsection{A mosaic permutation test}\label{subsec::mpt}

We seek to test the cluster-independence null $\mcH_0$ which asserts that $\{\epsilon_{C_m}\}_{m=1}^M$ are jointly independent. We now introduce a mosaic permutation test of $\mcH_0$ which is valid under Assumption \ref{assump::marginv} (marginal invariance).

The main idea is that marginal invariance plus cluster independence imply that the \textit{true} errors obey joint invariance (Assumption \ref{assump::jointinv}). However, naive estimated residuals based on a standard OLS regression will not obey Assumption \ref{assump::jointinv}. Instead, we will construct a \textit{mosaic} residual estimator which exactly obeys Assumption \ref{assump::jointinv}. 
Indeed, we will (a) use an \textit{invariance-augmented model} to ensure the estimated residuals are marginally invariant and (b) estimate residuals cluster-by-cluster to ensure the estimated residuals are cluster-independent. Together, these techniques ensure $\hat\epsilon$ satisfies joint invariance (Assumption \ref{assump::jointinv}). Formally, the method is as follows:

\begin{enumerate}[leftmargin=*, itemsep=0.5pt, topsep=0pt]
    \item \textit{Invariance-augmented model}: let $\bX^{(d)} \in \R^{N \times T}$ denote the matrix of values of the $d$th covariate. Define the \textit{$d$th transformed covariate} $\bX^{(\mathrm{trans},d)} \defeq \bX^{(d)} P$ by applying $P$ to $\bX^{(d)}$. Lastly, for each observation $i \in [N], t \in [T]$, define the augmented covariates $X\aug_{i,t} = (X^{(\mathrm{trans},d)}_{i,t})_{d=1}^D \in \R^{D}$ as the collection of transformed covariates for that observation.
    
    With this notation, we define the invariance-augmented model:
    \begin{equation}\label{eq::augmented_model}
        Y_{i,t} = \underbrace{{X_i,t}\trans \beta}_{\text{original model}} + \underbrace{{X_{i,t}\aug}^{\top} \beta\aug}_{\text{augmented component}} + \epsilon_{i,t}.
    \end{equation}
    This model is identical to the original model except that we have added $D$ additional ``transformed" covariates.
    \item \textit{Cluster-by-cluster estimation of residuals}:
    let $\hat\epsilon_{C_m}$ be the OLS residuals for the $m$th cluster based purely on the data from the $m$th cluster---i.e., the coefficients $\beta_d, \beta_d^{\mathrm{trans}}$ are estimated based only on $Y_{C_m}$. Let $\hat\epsilon \in \R^{N \times T}$ denote the appropriate concatenation of these residuals. Following \cite{mosaic_factor_paper2024}, we refer to these as ``mosaic" residuals, since the cluster-by-cluster estimation is reminiscent of a tiling in a mosaic. We define the notation $\hat\epsilon = \MosaicResid(Y, \bX)$ to emphasize that $\hat\epsilon$ is a function of $Y$ and all the covariates $\bX = (\bX^{(d)})_{d=1}^D$. 
    \item \textit{Randomization}: Form a randomized matrix of residuals $\tilde\epsilon \in \R^{N \times T}$ by sampling $B_1, \dots, B_M \iid \Bern(0.5)$
    and defining
    \begin{equation}
        \tilde\epsilon_{C_m} = B_m \hat\epsilon_{C_m} + (1 - B_m) \hat\epsilon_{C_m} P = \hat\epsilon_{C_m} P^{B_m}.
    \end{equation}
    In words, $\tilde\epsilon$ is formed by separately and independently transforming each cluster of residuals with $50\%$ probability, and otherwise leaving it unchanged. As notation, we let the function $\tilde\epsilon = \MosaicRandomize(\hat\epsilon)$ compactly denote this whole (randomized) process. Repeating this process $r=1, \dots, R$ times yields randomized residual matrices $\tilde\epsilon^{(1)}, \dots, \tilde\epsilon^{(R)}$.
    \item \textit{Test statistic}: for any test statistic $S : \R^{N \times T} \to \R$, the following is a valid p-value to test the null $\mcH_0$:
    \begin{equation}
        \pval \defeq \frac{1 + \sum_{r=1}^R \I(S(\hat\epsilon) \le S(\tilde\epsilon^{(r)}))}{R+1}.
    \end{equation}
    $S$ can be any test statistic meant to quantify evidence against the null---e.g., the average absolute correlation between the time series of residuals for units in different clusters. In Section \ref{subsec::robust_test}, we will introduce a default choice. 
\end{enumerate}

\begin{theorem}[Finite-sample validity of the test]\label{thm::test_finite_sample} Suppose the true errors satisfy Assumption \ref{assump::marginv}. Then under the null $\mcH_0$ that $\{\epsilon_{C_m}\}_{m=1}^M$ are independent,
\begin{equation}
    \P(\pval \le \alpha) \le \alpha \text{ for all } \alpha \in (0,1).
\end{equation}
This holds regardless of the choice of test statistic $S(\cdot)$.
\end{theorem}

Our method builds on many existing methods from the literature; indeed, several components of our test have appeared in different contexts. We discuss this below.

\begin{remark}[Augmented model]\label{rem::augment_model}
\cite{guan2024palmrt, mosaic_factor_paper2024} also use augmented models similar to Eq. \eqref{eq::augmented_model} to develop finite-sample valid permutation tests for homoskedastic regression and fundamental factor models, respectively. However, we use these augmented models in a different way. First, unlike \cite{guan2024palmrt}, we use cluster-by-cluster estimation, which (a) ensures the clusters of estimated residuals remain independent and (b) avoids the need to refit the entire linear model $R$ times, which is computationally prohibitive in the panel data setting. Second, unlike \cite{mosaic_factor_paper2024}, our test can pool information across multiple time periods. Lastly, a key difference relative to all existing methods is the precise form of the marginal and joint invariance assumptions. This allows us to show that our test is asymptotically robust to near-arbitrary violations of the invariance assumption, which would not be true if our method exploited a larger subgroup of invariances.
\end{remark}

\begin{remark}[Cluster-by-cluster estimation] Following several existing tests of cluster independence \citep[e.g.,][]{ibragimov2016}, we separately estimate residuals within each cluster. Unlike existing tests, we do not require that any particular regression coefficient can be estimated within each cluster---e.g., if $\bX^{(d)} = 0$ for some $d$, this does not pose a problem for our test---but we do require that the clusters are large enough to estimate nontrivial (i.e., nonzero) residuals. A standard way to ensure this is to ``combine" the smaller clusters (e.g., counties) into larger clusters (e.g., states).
\end{remark}

\begin{remark}[Finite-sample validity has a small cost] One might worry that these techniques greatly reduce the effective sample size in each regression, yielding low-quality residual estimates $\hat\epsilon$. For example, if we split the data into $20$ clusters and then double the number of covariates, this appears to reduce the effective sample size by a factor of $40$. However, this simple calculation is misleading, since in panel data, the dimensionality $D$ is often driven primarily by unit-specific fixed effects. E.g., imagine we have $N \approx 1000$ units, $T \approx 10$ time periods, and we run a regression with $D_0 \approx 20$ controls plus two-way fixed effects. Thus, we have $1000$ unit-specific fixed effects and $\approx 30$ other covariates. The next two paragraphs show that the price we pay for (i) augmenting the model and (ii) cluster-by-cluster estimation is driven by the small number ($30$) of other covariates, not by the large number (1000) of unit-specific fixed effects.
\begin{itemize}[itemsep=0.5pt, topsep=0pt, leftmargin=*]
    \item \textit{Augmenting the model}: When using any of the invariances in Section \ref{sec::invariances}, all of the ``augmented" covariates produced by the unit-specific fixed effects will be colinear with the original unit-specific fixed effects, and thus can be dropped. Thus, augmenting the model increases $D$ from $\approx 1030$ to only $\approx 1060$. 
    \item \textit{Cluster-by-cluster estimation}: inaccuracies in estimating the errors $\epsilon$ are largely driven by inaccuracies in estimating the unit-specific fixed effects. However, the observations from unit $i$ contain no information about the fixed effect for unit $j$. Therefore, when estimating residuals cluster-by-cluster, we obtain unit-specific fixed effect estimates---and therefore residual estimates---which are nearly as good as if we had pooled information from all clusters.
\end{itemize}
In short, in a typical panel data context, augmenting the model and estimating residuals cluster-by-cluster should not substantially reduce the quality of the estimated residuals.
\end{remark}

\subsection{Robustness to violations of the invariance assumption}\label{subsec::robust_test}

Theorem \ref{thm::test_finite_sample} shows our method is an exactly valid test of the cluster-independence null $\mcH_0$ under Assumption \ref{assump::marginv}. We now show that our test, when using a natural quadratic test statistic, asymptotically controls the Type I error rate under classical assumptions (Assumption \ref{assump::conventional}) even when the marginal invariance assumption is arbitrarily inaccurate. 

Our approach does not require studentizing or prepivoting the test statistic as in, e.g., \cite{neuhaus1993, janssen_thorsten_2005, chung_romano_2013, chung_romano_2016_multivariate, chung_romano_2016_ustats}, which is important since it is not clear how to estimate $\var(S(\hat\epsilon))$. A further challenge compared to previous work is that the asymptotic law of $S$ is non-universal and depends on the underlying distribution of $\epsilon$ \citep{bhattacharya2022asymptoticdistributionrandomquadratic}. Despite this, the precise form of Assumption \ref{assump::marginv} allows our test to automatically recover this non-universal limiting distribution.

We proceed in four parts: notation and assumptions (Section \ref{subsec::key_assumptions}), the key technical result (Section \ref{subsec::finite_sample}), the main result on Type I error control (Section \ref{subsec::asymptotic}), and numerical illustrations (Section \ref{subsubsec::test_robustness_sims}).

\subsubsection{Key assumptions and notation}\label{subsec::key_assumptions}

We now introduce notation and assumptions for our robustness results. Throughout, we consider test statistics that are a weighted sum of inter-cluster residual covariances:
\begin{equation}\label{eq::test_stat}
    S(\hat\epsilon) \defeq \sum_{m=1}^M \sum_{i \in C_m} \sum_{j \not \in C_m} s_{ij} \hat\epsilon_i\trans \hat\epsilon_j.
\end{equation}
Essentially, $S(\hat\epsilon)$ equals the weighted sum of the empirical covariances between the time-series of residuals $\hat\epsilon_i$ and $\hat\epsilon_j$ for units in different clusters, in which $\{s_{ij}\}_{i,j \in [N]}$ specifies the weights. 
To maximize power, we should choose weights such that $s_{ij} > 0$ if and only if we expect $\hat\epsilon_i$ to be correlated with $\hat\epsilon_j$.
We require throughout that the weights are normalized such that $\sum_{i \in C_m} \sum_{j \in C_{m'}} s_{ij}^2 = 1$.

\begin{equation}\label{eq::Delta_def}
    \Delta \defeq \frac{S - \tilde\E[\tilde S]}{\sigma} \text{ and } \tilde\Delta \defeq \frac{\tilde S - \tilde\E[\tilde S]}{\sigma}
\end{equation}
where $\sigma = \sqrt{\var\left(S - \tilde\E[\tilde S]\right)}$ is chosen such that $\var(\Delta) = 1$.   Conditional on the data, $\Delta, \tilde\Delta$ are a deterministic monotone transformation of $S$ and $\tilde S$; therefore it suffices to show $\P(S > \tilde Q_{1-\alpha}(\tilde S)) = \P(\Delta > \tilde Q_{1-\alpha}(\tilde\Delta)) \approx \alpha$. 
To do this, we need two assumptions and one piece of notation.

\begin{assumption}\label{assump::meanzero} The errors are mean zero, i.e., $\E[\epsilon] = 0$.
\end{assumption}

\begin{assumption}\label{assump::subgaussian} The errors $\epsilon$ are jointly sub-Gaussian and therefore have finite Luxemburg norm $\sgconstant$.\footnote{The Luxemburg norm is defined as
    $\sgconstant = \inf\left\{K > 0 : \sup_{A \in \R^{T \times N} : \|A\|_{\mathrm{Fr}} = 1} \E\left[\exp\left(\tr(A\trans \epsilon)^2 / K^2 \right)\right] \le 2\right\},$ where $\|\cdot\|_{\mathrm{Fr}}$ denotes the Frobenius norm.
} \end{assumption}

\begin{remark} We assume $\epsilon$ is sub-Gaussian for convenience, but Proposition \ref{prop::finitesamplemoment} holds as long as all moments of $\epsilon$ exist. See  Appendix \ref{appendix::finitesamplemoment}, Remark \ref{rem::subg} for details.
\end{remark}

Lastly, notation: for distinct clusters $m \ne m' \in [M]$, define
\begin{equation}\label{eq::delta_def}
    \delta_{m, m'} \defeq  
    \sum_{i \in C_m, j \in C_{m'}} s_{ij} \hat\epsilon_i\trans (I_T - P) \hat\epsilon_j
\end{equation}
to be the change in the empirical covariance of $\hat\epsilon_i, \hat\epsilon_j$ before and after applying $P$ to $\hat\epsilon_j$ (or equivalently, since $P$ is symmetric, $\hat\epsilon_i$), averaged over pairs of units in cluster $m$ and $m'$. Define $\sigma_{\delta}^2$ to equal the scaled average variance among the  $\delta_{m,m'}$:
\begin{equation}\label{eq::sigma_delta_def}
    \sigma_{\delta}^2 \defeq \frac{1}{\binom{M}{2}} \sum_{m < m' \in [M]} \var\left(\frac{1}{\sqrt{T}} \delta_{m, m'}\right).
\end{equation}
When we do asymptotic analysis, we will require that $\sigma_{\delta}^2$ is bounded away from zero to avoid technicalities where, e.g., $\delta_{m,m'} = 0$ a.s. for most $m$.

\subsubsection{Key technical ingredient: finite-sample moment bounds}\label{subsec::finite_sample}

We now state our key technical result: a finite-sample bound showing that moments of the randomization distribution approximate the unconditional moments of $\Delta$, i.e., $\tilde\E[\tilde\Delta^K] \approx \E[\Delta^K]$ for all $K \in \N$.

\begin{proposition}\label{prop::finitesamplemoment} Let $\mu_K = \E[\Delta^K]$ and $\tilde\mu_K = \tilde\E[{\tilde\Delta}^K]$ denote the $K$th moments of the unconditional and randomization distributions. Under Assumptions \ref{assump::meanzero}-\ref{assump::subgaussian} and $\mcH_0$, for all $K \in \N$, there exists a constant $a_K$ depending only on $K$ and $\sgconstant$ such that the following holds for all $M$:
\begin{equation}
    \E[(\mu_K - \tilde\mu_K)^2] \le \frac{a_K}{\sigma_{\delta}^{2K}} \cdot \frac{1}{M}.
\end{equation}
\end{proposition}

Since $a_K$ does not change with $M$, Proposition \ref{prop::finitesamplemoment} suggests that as the number of clusters $M \to \infty$, all moments of the randomization distribution converge to the unconditional moments of the normalized test statistic. Furthermore, Lemma \ref{lem::Delta_subexp} shows that $\Delta$ is sub-exponential and thus its moments determine its distribution. This suggests that for large $M$, $\P(\Delta > \tilde Q_{1-\alpha}(\tilde\Delta)) \approx \P(\Delta > Q_{1-\alpha}(\Delta)) \approx \alpha$. Numerical experiments in Section \ref{subsubsec::test_robustness_sims} suggest that this approximate equality can be accurate when $M \approx 20$. We now prove this formally in the next section.

\subsubsection{Asymptotic Type I error control}\label{subsec::asymptotic}

To show formal asymptotic validity, we consider an arbitrary triangular array of datasets $\{Y\upM\}_{M=1}^{\infty}$. The sizes of the clusters, the dimensions and law of $Y\upM$, the fixed design matrix, and even the invariances in Assumption \ref{assump::marginv} may change arbitrarily with $M$.\footnote{To avoid measurability concerns, we do require that all random variables are defined on the same probability space.}

As notation, let $S_M \defeq S(\hat\epsilon\upM)$ and $\tilde S_M \defeq S(\tilde\epsilon\upM)$ denote the test statistic and permuted test statistic for the $M$th dataset in the triangular array, and let $\Delta_M$ and $\tilde\Delta_M$ be their normalized variants as in Eq. \eqref{eq::Delta_def}. Let $\sigma_{\delta,M}^2$ be the value of $\sigma_{\delta}^2$ from Eq. \eqref{eq::sigma_delta_def} for the $M$th dataset in the triangular array. 

We now assume two regularity conditions. First, we require that $\sigma_{\delta,M}^2$---the average variance of $\delta_{m,m'}$ averaged over all pairs of clusters---is bounded away from zero as $M \to \infty$. This rules out pathological examples where all but finitely many of the cluster errors have zero variance.

\begin{assumption}\label{assump::minsigma} $\liminf_{M \to \infty} \sigma_{\delta,M}^2 > 0$.
\end{assumption}

Second, we require that asymptotically, the CDF $F_M$ of $\Delta_M$ does not have any point masses in a quantitative sense specified below. This continuity condition is needed to formally show that if the randomization distribution converges to the unconditional law of $\Delta_M$, then the Type I error rate is close to $\alpha$.

\begin{assumption}\label{assump::anticoncentration} For some $\xi > 0$, $\limsup_{M \to \infty} |F_M(b) - F_M(a)| \le \xi |a - b|$ for all $a, b \in \R$.
\end{assumption}

Under these additional assumptions, our test asymptotically controls the Type I error rate.

\begin{theorem}\label{thm::asymptotic_result} 
Suppose Assumption \ref{assump::meanzero} (mean-zero errors), Assumption \ref{assump::subgaussian} (sub-Gaussianity) and the null $\mcH_0$ hold for every dataset $Y\upM$ in the triangular array. 
Then under Assumptions \ref{assump::minsigma}-\ref{assump::anticoncentration}, we have
\begin{equation}\label{eq::main_asymptotic_result}
    \limsup_{M \to \infty} \P(S_M > \tilde Q_{1-\alpha}(\tilde S_M)) \le \alpha.
\end{equation}
Furthermore, if $\lim_{M \to \infty} \P(S_M = \tilde Q_{1-\alpha}(\tilde S_M)) = 0$, then Eq. \eqref{eq::main_asymptotic_result} holds with equality.
\end{theorem}

\subsubsection{A numerical illustration of robustness}\label{subsubsec::test_robustness_sims}

We now perform a simulation with $N=200$ units, $T \in \{10,50\}$ observations per subject, and $M \in \{20, 40, 100, 200\}$ evenly sized clusters. We sample the errors as
\begin{equation}
    \epsilon_{i,t} = \rho \epsilon_{i,t-1} + t^{1/4} \sqrt{1-\rho^2} \gamma_{it} + \eta_{c(i),t}
\end{equation}
where $\gamma_{i,t}$ are i.i.d.~Laplace random variables with PDF $p(x) = \frac{1}{2} \exp(-|x|)$, $c(i)$ denotes the cluster of subject $i$, and $\eta_{c(i),t} \iid \mcN(0,1)$ represents a common cluster-specific error component. Overall, this ensures that the errors are auto-correlated (across time), non-stationary (since $\var(\epsilon_{i,t}) \approx 1 + \sqrt{t})$), and correlated within clusters. For simplicity, there is a single covariate drawn as an i.i.d.~Gaussian random variable.

\begin{figure}[!h]
    \centering
    \includegraphics[width=0.9\linewidth]{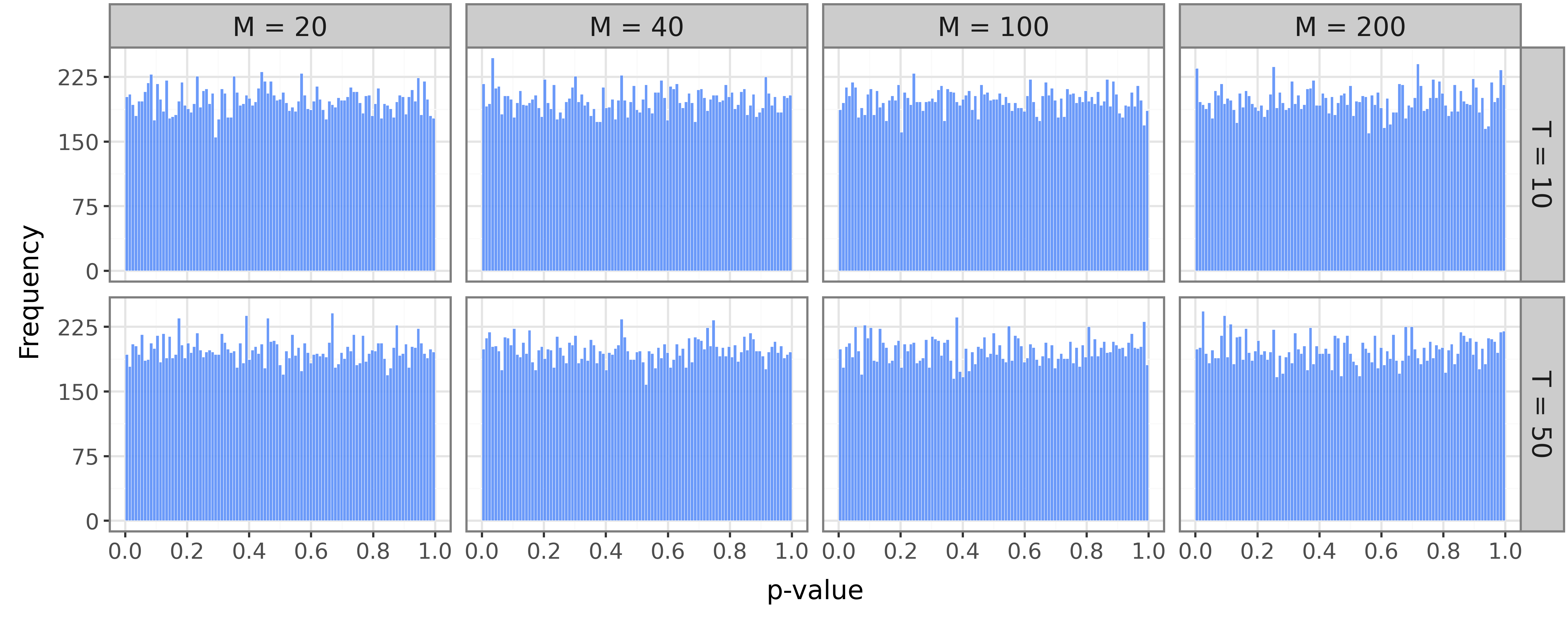}
    \caption{This figure shows the marginal distribution of mosaic p-values in the simulations from Section \ref{subsubsec::test_robustness_sims}.}\label{fig::pval}
\end{figure}

\begin{figure}[!h]
    \centering
    \includegraphics[width=\linewidth]{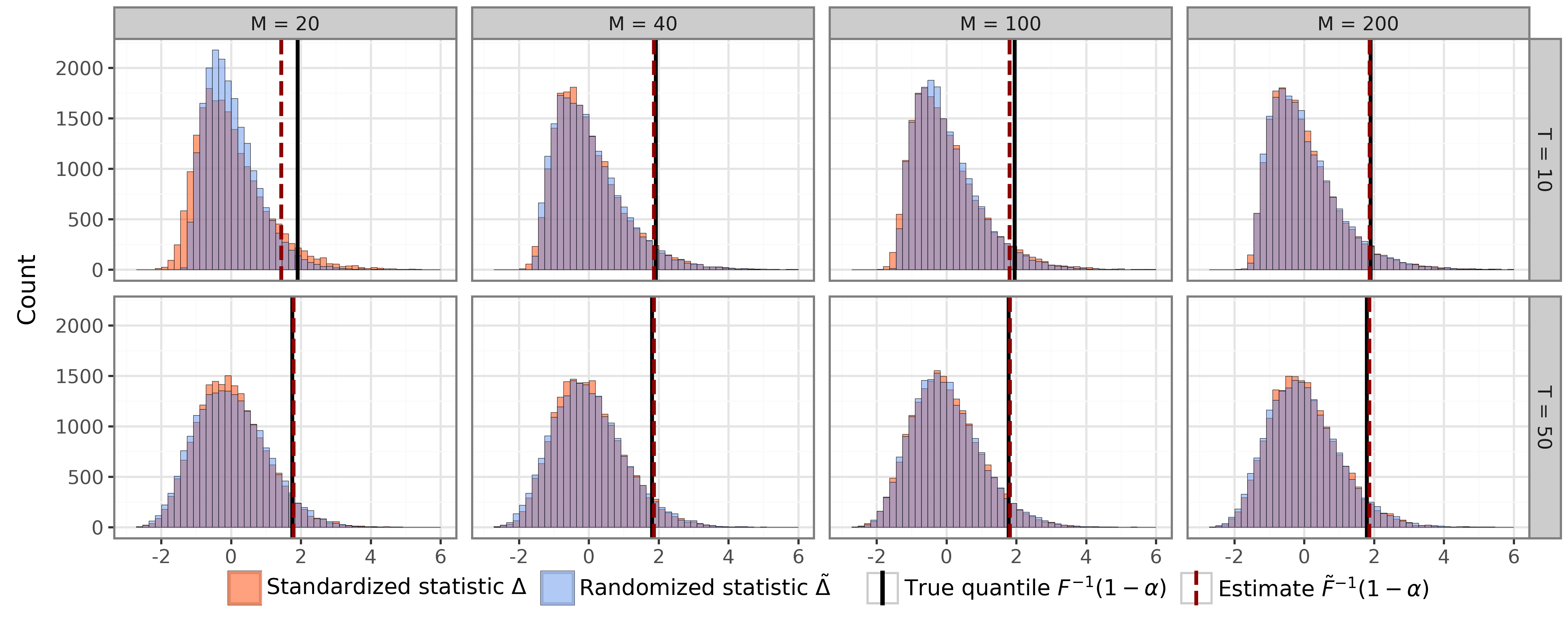}
    \caption{These histograms show the marginal law $F$ of the test statistic $\Delta$ and the randomization distribution $\tilde F$, i.e., the law of $\tilde\Delta$ conditional on the data. These results must be taken with a grain of salt since we only show the law of $\tilde\Delta$ conditional on one fixed dataset, so the results can depend slightly on the random seed. Nonetheless, we see that $F$ and $\tilde F$ are similar even though the shape of $F$ is non-universal. The black and dotted red lines show the $1-\alpha$ quantiles of $F$ and $\tilde F$, respectively.}\label{fig::Delta}
\end{figure}

We apply the mosaic permutation test with the ``local exchangeability" invariance assumption from Section \ref{subsec::mpt}. Although local exchangeability does not hold (due to autocorrelation and time-varying heteroskedasticity), Figure \ref{fig::pval} shows that our test produces approximately uniform p-values even with $M=20$ clusters. Next, Figure \ref{fig::Delta} shows that the randomization distribution accurately approximates the law of $\Delta$ even when $M$ and $T$ are small. It also shows that the law of $\Delta$ is not Gaussian and non-universal, since when $T=10$ the distribution is highly skewed, and for $T=50$ the distribution is more symmetric (although still somewhat skewed). Despite this, our test appears to accurately estimate the quantiles of $\Delta$.

\newpage
\section{Inference on linear models}\label{sec::cis}


We now show how to invert the test developed in Section \ref{sec::mpt} to obtain confidence intervals (CIs) for regression coefficients. As notation, suppose we have the following regression model:
\begin{equation}\label{eq::ci_model}
    Y_{i,t} = Z_{i,t} \beta\opt + X_{i,t}\trans \gamma + \epsilon_{i,t},
\end{equation}
where $Z_{i,t} \in \R$ is the covariate of interest and $X_{i,t} \in \R^D$ are other controls. We let $Z \in \R^{N \times T}$ be the matrix of covariate values. 
Under Assumption \ref{assump::jointinv} (joint invariance), we may invert a mosaic permutation test to obtain a confidence interval for $\beta$. That is, for each $b \in \R$, we define $Y_{i,t}(b) = Y_{i,t} - Z_{i,t} b$, and then we apply a mosaic permutation test on $Y_{i,t}(b)$ and covariates $X_{i,t}$, yielding a p-value $\pval(b)$. The final confidence interval is $\CI \defeq \{b \in \R : \pval(b) \ge \alpha \}.$

Under Assumption \ref{assump::jointinv}, this procedure yields finite-sample valid confidence intervals using any test statistic, as in Theorem \ref{thm::test_finite_sample}. However, we recommend using the following test statistic to ensure the CIs are (i) efficient to compute and (ii) robust to violations of the invariance assumption. As notation, define
\begin{equation}\label{eq::ci_mosaicresid}
    \hat\epsilon = \MosaicResid(Y, \bX) \text{ and } A = \MosaicResid(Z, \bX).
\end{equation}
In words, we form $A \in \R^{N \times T}$ by applying the mosaic residual function to the \textit{covariate} $Z$ instead of the outcome $Y$. Then we define $D \defeq \frac{1}{2} (A - A P)$ to be the scaled difference between the covariate residuals $A$ and their transformation under the invariance from Assumption \ref{assump::marginv}. Finally, we define the test statistic:
\begin{equation}\label{eq::hatbeta_mosaic_def}
    \hat\beta\mosaic = S(\hat\epsilon) \defeq \frac{\langle D, \hat\epsilon \rangle}{\langle D, D \rangle},
\end{equation}
where $\langle A, B \rangle \defeq \sum_{i,t} A_{i,t} B_{i,t}$. Intuitively, $\hat\beta\mosaic$ represents a ``mosaic" estimator of $\beta$, since if we replaced $D$ and $\hat\epsilon$ with the OLS residuals of $Z$ and $Y$ with respect to $\bX$, then $S(\hat\epsilon)$ would equal the OLS estimate of $\beta$. 

Using this test statistic, it is easy to invert our test, yielding transparent and computationally efficient confidence intervals. In particular, let $\tilde D = \MosaicRandomize(D)$ and $\tilde\epsilon = \MosaicRandomize(\hat\epsilon)$ denote the matrices $D, \hat\epsilon$ after the randomized transformation introduced in Section \ref{subsec::mpt}. Let $\tilde\beta = S(\tilde\epsilon)$ denote the test statistic applied to the randomized residuals. Finally, define $\slope$ to be the (random) angle between $D$ and $\tilde D$: 
\begin{equation}
    \slope \defeq \frac{\langle D, \tilde D \rangle}{\sqrt{\langle D, D \rangle\langle \tilde D, \tilde D \rangle}}.
\end{equation}
Then, if $\tilde Q_{\gamma}$ denotes a $\gamma$ quantile conditional on the data, the mosaic confidence interval is defined as follows:
\begin{equation}
    \mathrm{CI}\mosaic \defeq \left[\hat\beta\mosaic + \tilde Q_{\alpha/2}\left(\frac{\slope \hat\beta\mosaic - \tilde\beta}{1-\slope}\right), \hat\beta\mosaic + \tilde Q_{1-\alpha/2}\left(\frac{\slope \hat\beta\mosaic - \tilde\beta}{1-\slope}\right) \right].
\end{equation}

Why is $\CI\mosaic$ based off the quantiles of $\frac{\slope \hat\beta\mosaic - \tilde\beta}{1-\slope}$ instead of the quantiles of (e.g.) $\hat\beta\mosaic - \tilde\beta$? We interpret the numerator and denominator separately.
\begin{itemize}[itemsep=0.5pt, topsep=0pt, leftmargin=*]
    \item The numerator ensures the confidence interval is \textit{shift equivariant}, meaning that replacing $Y_{i,t}$ with $Y_{i,t} + b Z_{i,t}$ shifts the entire confidence interval by $b$. Another interpretation is that when the true errors are mean zero, $\tilde\rho$ is the unique function of $B_1, \dots, B_M$ such that $\E[\slope \hat\beta\mosaic - \tilde\beta \mid B_1, \dots, B_M] = 0$ holds in finite samples (see Appendix \ref{appendix::cis}, Remark \ref{rem::unbiased_bs_terms}).
    \item The denominator $1-\slope$ improves performance (intuitively) because as $\slope$ increases, we should expect $\hat\beta\mosaic \approx \tilde\beta$ to be more similar since $D \approx \tilde D$. Thus, as $\slope$ gets larger, any differences between $\hat\beta\mosaic$ and $\tilde\beta$ become stronger evidence of noise and should be magnified to widen the confidence interval.
    \end{itemize}

Theorem \ref{thm::ci_validity} confirms that $\CI\mosaic$ is a finite-sample valid confidence interval for $\beta$ assuming only joint invariance. The proof shows that $\CI\mosaic$ equals the interval obtained after inverting a mosaic permutation test using the test statistic from Eq. \eqref{eq::hatbeta_mosaic_def}.
\begin{theorem}\label{thm::ci_validity} Suppose that Assumption \ref{assump::jointinv} holds. Then $\P(\beta\opt \in \CI_{\mosaic}) \ge 1 - \alpha$.
\end{theorem}

\textbf{Robustness}. $\CI\mosaic$ is asymptotically valid under standard asymptotics (Assumption \ref{assump::conventional}), even when joint invariance fails. To show this, we follow Section \ref{subsec::asymptotic} and consider a triangular array of datasets $\{Y\upM\}_{M=1}^{\infty}$ with covariates $\{Z\upM\}_{M=1}^{\infty}$ and controls $\{\bX\upM\}_{M=1}^{\infty}$. We make no assumptions about the sequence of datasets except to assume (1) mean zero errors (Assumption \ref{assump::meanzero}), (2) cluster independence, and (3) the following Lyapunov condition.

\begin{assumption}\label{assump::lyapunov} Let $\hat\epsilon = \MosaicResid(Y\upM - \beta\opt Z\upM, \bX\upM)$ denote oracle mosaic residuals formed after subtracting off the true influence of the covariate $Z\upM$, and let $A = \MosaicResid(Z\upM, \bX\upM)$ be the mosaic-orthogonalized covariate, and $D = A - A P$. Define $\hat\theta_m = \langle D_{C_m}, \hat\epsilon_{C_m} \rangle - \langle D_{C_m}, \hat\epsilon_{C_m} P \rangle$, so $\hat\theta_m$ measures the difference in the inner product between $D_{C_m}$ and $\hat\epsilon_{C_m}$ before and after the residuals are transformed by $P$. We assume that as $m \to \infty$,
\begin{equation}
    \frac{\sum_{m \in [M]}\E[|\hat\theta_m|^{2+\delta}]}{\sigma_M^{2+\delta}} \to 0 \text{ where } \sigma_M^2 = \sum_{m \in [M]} \var(\hat\theta_m).
\end{equation}
Furthermore, for some $\delta_1 > \delta$, $\E[|\hat\theta_m|^{2+\delta_1}|]$ is uniformly bounded and $\var(\hat\theta_m)$ is bounded from below. 
\end{assumption}

Assumption \ref{assump::lyapunov} essentially requires that no individual cluster produces a cluster-level estimate $\hat\theta_m$ that dominates the overall test statistic. Note that the minimum variance condition on $\var(\hat\theta_m)$ requires that there is variation among the covariate of interest within each cluster; this can be automatically satisfied by merging clusters until this is true. There is no disadvantage to merging in this fashion, since whenever $D_{C_m} = 0$, cluster $m$ does not affect the confidence interval in any way.

\begin{theorem}[Robust CIs]\label{thm::robust_cis} Under Assumption \ref{assump::meanzero} (mean-zero errors), cluster independence (Eq. \ref{eq::null}), and Assumption \ref{assump::lyapunov} (Lyapunov condition),  $\liminf_{M \to \infty} \P(\beta\opt \in \CI\mosaic\upM) \ge 1 - \alpha.$
\end{theorem}

Together, Theorem \ref{thm::ci_validity} and Theorem \ref{thm::robust_cis} show that $\CI\mosaic$ is valid under arguably weaker assumptions than many classical methods, since $\CI\mosaic$ is asymptotically valid under classical asymptotics and finite-sample valid under joint invariance.

\begin{remark}[Standard errors] $\CI\mosaic$ is not based on a standard error estimate. However, a natural standard error estimator is below:
\begin{equation}
    \hat\sigma\mosaic \defeq \widetilde{\mathrm{SD}}\left(\frac{\slope \hat\beta\mosaic - \tilde\beta}{1 - \slope}\right),
\end{equation}
where the standard deviation is taken conditional on the data. The standard error estimate $\hat\sigma_{\mosaic}$ is the analog of the sample quantiles taken in the definition of $\CI\mosaic$.
\end{remark}

\begin{remark}[Length] Mosaic confidence intervals can either be narrower or wider than (e.g.) typical cluster-robust OLS intervals. For example, mosaic confidence intervals based on local exchangeability will typically be narrower than OLS ones if the errors are highly autocorrelated. On the other hand, if the errors are i.i.d.~Gaussian, OLS intervals will be narrower. In experiments in Section \ref{sec::experiments}, we find that mosaic confidence intervals are typically $10$-$50\%$ wider than existing intervals, but they appear to more accurately quantify uncertainty.
\end{remark}

\section{Empirical diagnostics on real data}\label{sec::experiments}

We now run empirical tests to diagnose which methods best quantify uncertainty for linear models of panel data. We analyze three datasets from \cite{vella_verbeeck1998_unions, gentzkow2011newspaper, cao2022aer}, each of which has been well-studied in the literature. All code, data, and results are publicly available on Github at \url{https://github.com/amspector100/mosaic_panel_paper}.

On these datasets, we evaluate five methods: naive OLS assuming homoskedasticity, OLS with cluster-robust errors, HAC errors, the wild residualized cluster (WCR) bootstrap \citep{mackinnon2023clusterrobust}, and mosaic confidence intervals. We choose the data preprocessing steps, the clustering $\{C_m\}_{m=1}^M$, and the choice of controls in each analysis to match the original papers. For example, \cite{gentzkow2011newspaper} take first differences before estimating the model, as do we. For \cite{vella_verbeeck1998_unions, gentzkow2011newspaper}, we reuse preprocessing code from \cite{twfe2020aer}.

In each dataset, we perform two empirical diagnostics. Most existing methods produce an estimator $\hat\beta$ and a standard error $\hat\sigma$ such that $\hat\beta$ is asymptotically normal and $\hat\beta \pm \Phi^{-1}(\alpha/2) \hatse$ is an asymptotically valid confidence interval: 
\begin{equation}\label{eq::asymp_normal}
    \frac{\hat\beta- \beta\opt}{\hatse} \to \mcN(0,1).
\end{equation}
To diagnose this result, we split the data into two disjoint folds by splitting the clusters into two disjoint groups. This splitting can be done \textit{after} looking at the design matrix (since the design matrix is fixed). In our analysis, we create the two folds by (a) randomly selecting one cluster and (b) letting the first fold be formed by the set of $\floor{M/2}$ clusters that are closest (in space) to the randomly selected cluster. The second fold is formed by the other $\ceil{M/2}$ clusters. The exception is \cite{vella_verbeeck1998_unions}, where units do not have an associated location in space; in this case, we run the same procedure but treat the pre-treatment covariates as ``coordinates."

Then for $k \in \{1,2\}$, we fit estimators, standard errors, and level $\alpha$ confidence intervals $\hat\beta_k, \hatse_k, \CI_{\alpha,k}$ on the $k$th fold of data. Assuming cluster independence, we have joint asymptotic normality:
\begin{equation}
    \left(\frac{\hat\beta_1- \beta\opt}{\hatse_1}, \frac{ \hat\beta_2- \beta\opt}{\hatse_2}\right) \tod \mcN(0, I_2).
\end{equation}
We then perform two empirical diagnostics. 

\textbf{Diagnostic 1: standard errors.} If the folds are independent, $\hat\beta_1, \hat\beta_2$ are unbiased, and the standard errors are accurate, then
\begin{equation}\label{eq::ratios}
    \Lambda \defeq \frac{\left(\hat\beta_1 - \hat\beta_2\right)^2}{\hatse_1^2 + \hatse_2^2} \text{ satisfies } \E[\Lambda] \approx 1.
\end{equation}

Figure \ref{fig::ratios} shows the empirical average of the ratio in the previous equation, averaged across many folds of data and the relevant features of interest in each dataset. (De-averaged results are available on GitHub.) 
Again, we find that the average ratios for the mosaic methods are much closer to one than the traditional OLS-based methods.

\textbf{Diagnostic 2: overlapping confidence intervals.} Since $\CI_{\alpha,1}$ and $\CI_{\alpha,2}$ are independent, asymptotically valid confidence intervals covering the same parameter, they should overlap with high probability. Formally, a standard calculation shows that under Eq. \eqref{eq::asymp_normal}, we have an asymptotically consistent estimator of the probability of overlap as follows:    \begin{equation}\label{eq::testable_implication}
    \underbrace{1 - 2 \Phi\left(\Phi^{-1}(\alpha/2)\sqrt{\frac{
            \left(\hatse_1 + \hatse_2\right)^2
    }{
        \hatse_1^2 + \hatse_2^2
    }}\right)}_{\defeq \hat p\theor(\alpha)} - \underbrace{\P(\CI_{\alpha,1} \cap \CI_{\alpha,2} \ne \emptyset)}_{p\emp(\alpha)} \toprob 0.
    \end{equation}

We refer to $\hat p\theor(\alpha)$ as the \textit{theoretical} probability of overlap since it relies on Eq. \ref{eq::asymp_normal} and cluster-independence. However, we can also estimate the \textit{empirical} probability of overlap $\hat p\emp(\alpha)$ by splitting the data many times and checking the proportion of times $\CI_{\alpha,1}$ and $\CI_{\alpha,2}$ overlap. After averaging over many random splits, if $\hat p\emp(\alpha)$ is smaller than $\hat p\theor(\alpha)$, this suggests that the original inference procedure is anticonservative.

Figure \ref{fig::coverage_results} plots $\hat p\theor(\alpha)$ against $\hat p\emp(\alpha)$. We find that the mosaic methods perform quite well, whereas the other methods seem anticonservative.

There are many possible explanations for the results in Figures \ref{fig::ratios} and \ref{fig::coverage_results}. For example, mosaic methods only rely on a joint invariance assumption (Assumption \ref{assump::jointinv}), which is neither weaker nor stronger than cluster independence; perhaps this assumption is more accurate than standard assumptions for the datasets we have analyzed (Assumption \ref{assump::conventional}). Another explanation is that mosaic methods based on time reversal and local exchangeability do not require the errors to be mean zero; perhaps this reduces bias and makes them robust to certain types of unobserved confounders. Indeed, the time reversal and local exchangeability assumptions appear to outperform the symmetry assumption in our experiments, giving partial support to this explanation. A last note is that all other methods only offer asymptotic guarantees, but in the context of panel data, asymptotic results can depend delicately on the asymptotic regime \citep[see, e.g.,][]{mackinnon2023clusterrobust}. Thus, the finite-sample guarantees of mosaic methods might also explain their performance. At present, we do not have a perfect explanation for these empirical results. However, they do give confidence that mosaic methods can improve uncertainty quantification in real empirical applications.

\begin{figure}[!h]
    \centering
    \includegraphics[width=\linewidth]{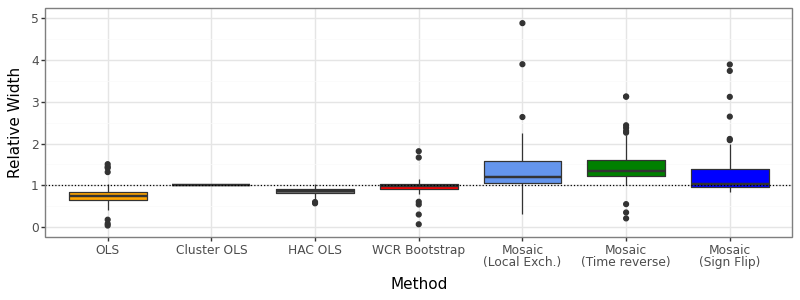}
    \caption{For each method, this figure shows the distribution of widths of confidence intervals relative to standard cluster OLS confidence intervals (see Section \ref{sec::experiments} for details). Note that for readability, this plot excludes three outliers where mosaic (local exch.) produces confidence intervals that are (on average) $8.5, 14.6$ and $81.6$ times wider than cluster OLS; these outliers are produced when analyzing covariates that have almost no local variation within units.}\label{fig::widths}
\end{figure}

\textbf{Width and power}: Empirically, mosaic methods produce wider confidence intervals than existing methods. For each dataset and feature of interest, we compute the median width of the confidence intervals produced by each method (averaged over many choices of the two folds). Then, we compute the ratio between these median widths and the median width of OLS cluster-robust intervals. Figure \ref{fig::widths} shows the distribution of these ratios. We find that mosaic methods confidence intervals are typically $1.1$ to $1.5$ times wider than existing methods. This is likely driven by two distinct phenomena, namely that (i) existing confidence intervals appear to be too narrow and (ii) mosaic estimators are on average higher variance than the standard OLS estimators (although this result depends on the setting, and mosaic estimators are sometimes lower variance). A $1.1$ to $1.5$ times increase in width is arguably a reasonable price to pay for improved uncertainty quantification. However, this result also motivates future work on methods that minimize width while accurately quantifying uncertainty in analyses of panel data.

\section{Discussion}

This paper introduces a mosaic permutation test for inference on linear models of panel data. Our test can be used for two purposes. First, it can test the cluster independence assumption. Second, it can produce confidence intervals for linear models. In both cases, our method is finite-sample valid under an invariance assumption, and it produces asymptotically valid results under the standard cluster-independence assumption (without any invariance assumption). Our paper introduces several practical choices of invariance assumptions, and we show in experiments on real datasets that our methods more accurately quantify uncertainty than several existing competitors.

However, our method has several important weaknesses that may motivate future work. 
\begin{itemize}[topsep=0pt, itemsep=0.5pt, leftmargin=*]
    \item \textit{Cluster-by-cluster estimation}: to obtain finite sample results, 
    our method requires the analyst to estimate residuals cluster-by-cluster. This may result in lower power, and if the clusters are too small, it also requires the analyst to make potentially arbitrary decisions about how to combine the clusters into a set of sufficiently large ``coarse" clusters. 
    
    \item \textit{Power against alternatives satisfying joint invariance}: our test of cluster independence is highly robust, but it is powerless against alternatives that satisfy joint invariance (Assumption \ref{assump::jointinv}). We can circumvent this by applying the test with several different invariance assumptions and applying a multiplicity correction (since asymptotically, each test will be valid under the cluster-independence null even if the marginal invariance assumption is false---see Theorem \ref{thm::asymptotic_result}), but nonetheless it would be interesting to find other ways to improve the power of the test. 
    
    \item \textit{Multi-way clustering}: our methods do not apply in situations where the analyst wishes to employ multi-way clustering. Extensions to this context could prove fruitful.

    \item \textit{Non-linear models}: our methods provide finite-sample inferential guarantees only for linear models (Eq. \ref{eq::linear_model}), 
    but it might be valuable to develop extensions to other models.

    \item \textit{Other invariances}: we give three examples of invariances (Assumption \ref{assump::marginv}). However, our method could use other invariances, which might improve empirical performance, for example by reducing the width of the confidence intervals.

    \item \textit{Asymptotics when $M$ is small}: our asymptotic robustness results require the number of clusters $M$ to diverge. However, \cite{CRS_2017, canay_permutation_tests2017} develop an approach to show robustness when $M$ is fixed, namely to show that an invariance assumption holds asymptotically. It would be interesting to investigate if our methods are robust in this asymptotic regime; if so, this would complement our existing results.
\end{itemize}

\bibliography{ref}
\bibliographystyle{apalike}

\appendix 

\section{Proofs for finite-sample validity results}

\subsection{Proof of Theorem \ref{thm::test_finite_sample}}

Before proving Theorem \ref{thm::test_finite_sample}, we prove a pair of simple lemmas. These lemmas essentially prove the main content of Theorem \ref{thm::test_finite_sample}. Note that for convenience, they use vectorized notation.

\begin{lemma}[Properties of the augmented projection matrix]\label{lem::aug_model} Let $\bX \in \R^{n \times d}$ and let $P \in \R^{n \times n}$ be a symmetric idempotent matrix. Define the augmented design $\tilde\bX \defeq [\bX, P \bX] \in \R^{n \times 2d}$ and the augmented projection matrix $H \defeq \tilde\bX \left(\tilde\bX\trans \tilde\bX\right)^{-1} \tilde\bX\trans$. Then $P H P = H$.
\begin{proof} Define $A \defeq P \tilde X = [P \bX, \bX]$ and $H_A \defeq A (A\trans A^{-1}) A\trans$. Note that $H = H_A$ since the OLS projection matrix does not depend on the order of the columns of the design matrix. However, we also have that
\begin{equation}\label{eq::HeqPHP}
    H = H_A = P \tilde X ((P\tilde X) \trans P\tilde X)^{-1} \tilde X\trans P\trans = P \tilde X (\tilde X\trans \tilde X)^{-1} \tilde X^T P = P H P,
\end{equation}
where the second to last equation follows because $P\trans P = P^2 = I_T$ and $P\trans = P$.
\end{proof}
\end{lemma}

\begin{lemma}[Mosaic residuals preserve joint invariance]\label{lem::vec_jointinv} Let $\bX \in \R^{n \times d}$, $\epsilon \in \R^n$, and for some $\beta \in \R^d$, $\by = \bX \beta + \epsilon$. Let $C_1, \dots, C_M \subset [n]$ partition $[n]$, and for each $m \in [M]$, let $P_m \in \R^{|C_m| \times |C_m|}$ be a symmetric idempotent matrix. For each $m \in [M]$, define the augmented design $\tilde\bX_{(m)} \defeq [\bX_{C_m}, P_m \bX_{C_m}] \in \R^{|C_m| \times 2d}$ and the augmented projection matrix $H_m \defeq \tilde\bX \left(\tilde\bX\trans \tilde\bX\right)^{-1} \tilde\bX\trans$. Define mosaic residuals as $\hat\epsilon_{C_m} \defeq H_m Y_{C_m} = H_m \epsilon_{C_m}$.

Suppose that for any $z_1, \dots, z_M \in \{0,1\}$, 
\begin{equation}\label{eq::jointinv_vec}
    (\epsilon_{C_1}, \dots, \epsilon_{C_M}) \disteq (P_1^{z_1} \epsilon_{C_1}, \dots, P_M^{z_M} \epsilon_{C_M}). 
\end{equation}
Then this result also holds for the mosaic residuals:
\begin{equation}
    (\hat\epsilon_{C_1}, \dots, \hat\epsilon_{C_M}) \disteq (P_1^{z_1} \hat\epsilon_{C_1} \dots, P_M^{z_M} \hat\epsilon_{C_M}).
\end{equation}
\begin{proof} By Lemma \ref{lem::aug_model}, we know that $P_m H_m P_m = H_m$ for all $m \in [M]$. This implies $P_m^{z_m} H_m P_m^{z_m} = H_m$ for all $m \in [M]$. Therefore, we have that
\begin{align*}
    (\hat\epsilon_{C_1}, \dots, \hat\epsilon_{C_M})
    &=
    (H_1 \epsilon_{C_1}, \dots, H_M \epsilon_{C_M}) & \text{ by definition} \\
    &=
    (P_1^{z_1} H_1 P_1^{z_1} \epsilon_{C_1}, \dots, P_M^{z_M} H_M P_M^{z_M} \epsilon_{C_M}) & \text{ since } P_m^{z_m} H_m P_m^{z_m} = H_m \\
    &\disteq
    (P_1^{z_1} H_1 \epsilon_{C_1}, \dots, P_M^{z_M} H_M \epsilon_{C_M}) & \text{ by Eq. \eqref{eq::jointinv_vec}} \\
    &=
    (P_1^{z_1} \hat\epsilon_{C_1}, \dots, P_M^{z_M} \hat \epsilon_{C_M}) & \text{ by definition.}
\end{align*}
\end{proof}
\end{lemma}

Now we prove Theorem \ref{thm::test_finite_sample}.

\begingroup
\def\thetheorem{\ref{thm::test_finite_sample}}
\begin{theorem} Suppose the true errors satisfy Assumption \ref{assump::marginv}. Then under the null $\mcH_0$ that $\{\epsilon_{C_m}\}_{m=1}^M$ are independent,
\begin{equation}
    \P(\pval \le \alpha) \le \alpha \text{ for all } \alpha \in (0,1).
\end{equation}
\begin{proof} The proof proceeds in two steps. Note that throughout the proof, we only use the fact that the true errors satisfy Assumption \ref{assump::jointinv}. This is implied by the null $\mcH_0$ plus Assumption \ref{assump::marginv}, but we do not otherwise use these assumptions.

\textbf{Step 1}: We first show that for any $z_1, \dots, z_M \in \{0,1\}$,
\begin{equation}
    (\hat\epsilon_{C_1}, \dots, \hat\epsilon_{C_M}) \disteq (\hat\epsilon_{C_1} P^{z_1}, \dots, \hat\epsilon_{C_M} P^{z_M}).
\end{equation}
To do this, we will apply Lemma \ref{lem::vec_jointinv}. However, this lemma is stated in terms of vectors of errors, not matrices of errors. To apply Lemma \ref{lem::vec_jointinv}, we note that this invariance assumption can be stated in a vectorized form:
\begin{align}
    (\hat\epsilon_{C_1}, \dots, \hat\epsilon_{C_M}) &\disteq (\hat\epsilon_{C_1} P^{z_1}, \dots, \hat\epsilon_{C_M} P^{z_M}) \label{eq::matrix_jointinv} \\
\text{ holds if and only if } 
    \left(\vecop(\hat\epsilon_{C_1}), \dots, \vecop(\hat\epsilon_{C_M})\right) &\disteq \left((P^{z_1} \otimes I_{|C_1|}) \vecop(\hat\epsilon_{C_1}), \dots, (P^{z_M} \otimes I_{|C_M|}) \vecop(\hat\epsilon_{C_M})\right). \label{eq::vec_jointinv}
\end{align}
where this follows from the standard fact that $\vecop(AB) = (B^T \otimes I_k) \vecop(A)$ for any $A \in \R^{k \times \ell}, B \in \R^{\ell \times m}$, as well as the fact that $P$ is symmetric by assumption. Above, $\otimes$ denotes the Kronecker product.

Since $P$ is symmetric and idempotent, $P_m \defeq P\trans \otimes I_{|C_m|}$ is also symmetric and idempotent. Furthermore, each cluster of empirical residuals $\hat\epsilon_{C_m}$ is formed by a cluster-specific OLS regression on the augmented design as specified in Lemma \ref{lem::vec_jointinv}. Thus, $\vecop(\hat\epsilon_{C_m})$ satisfies the assumptions of Lemma \ref{lem::vec_jointinv} and we conclude that Eq. \eqref{eq::vec_jointinv} holds, which implies Eq. \eqref{eq::matrix_jointinv} holds.



\textbf{Step 2}: at this point, standard statistical arguments show that if we sample $B_1, \dots, B_M \iid \Bern(0.5)$ and we define $\tilde\epsilon$ to be the appropriate concatenation of $(\hat\epsilon_{C_m} P_m^{B_m})_{m=1}^M$, then $(\hat\epsilon, \tilde\epsilon)$ are exchangeable. Furthermore, if we repeat this process $R$ times, yielding $R$ transformed matrices $\{\tilde\epsilon^{(1)}, \dots, \tilde\epsilon^{(R)}\}$, then $(\hat\epsilon, \tilde\epsilon^{(1)}, \dots, \tilde\epsilon^{(R)})$ are jointly exchangeable and $\P(\pval \le \alpha) \le \alpha$. In short, this follows from Eq. \eqref{eq::matrix_jointinv} plus the fact that the collection of transformations $(\hat \epsilon_{C_1}, \dots, \hat \epsilon_{C_M}) \mapsto (\hat\epsilon_{C_1} P^{z_1}, \dots, \hat\epsilon_{C_M} P^{z_M})$ form a subgroup. Please see, e.g., \cite{mosaic_factor_paper2024} for rigorous proofs of the arguments in Step 2. 
\end{proof}
\end{theorem}
\endgroup

\subsection{Proof of Theorem \ref{thm::ci_validity}}\label{appendix::cis}

The main idea behind Theorem \ref{thm::ci_validity} is to show that the mosaic confidence interval $\CI\mosaic$ defined in Section \ref{sec::cis} is an inversion of the mosaic permutation test from Section \ref{sec::mpt}. To do this, we need the following Lemma.

\begin{lemma}\label{lem::lm_fact} Fix any $\by \in \R^n, \bX \in \R^{n \times p}, \bz \in \R^n.$ Let $\hat\epsilon(b) \in \R^n$ denote the estimated OLS residuals when regressing $\by(b) \defeq \by - b \bz$ onto $\bX$ and let $\ba$ be the OLS residuals when regressing $\bz$ onto $\bX$. Then
\begin{equation}
    \hat\epsilon(b) = \hat\epsilon(0) - b \ba.
\end{equation}
\begin{proof} Let $H = \bX (\bX\trans \bX)^{-1} \bX\trans$ be the OLS projection matrix, so that $\ba = H \bz$. Furthermore,
\begin{equation*}
    \hat\epsilon(b) \defeq H \by(b) = H(\by(0) - b \bz) = H \by(0) - b H \bz = \hat\epsilon(0) - b \ba.
\end{equation*}
\end{proof}
\end{lemma}

Following the notation of Section \ref{sec::cis}, let $\hat\epsilon(b) = \MosaicResid(Y - b Z, \bX) \in \R^{N \times T}$ denote the mosaic residuals run on the shifted outcomes $Y - bZ$. Within each cluster, $\hat\epsilon(b)$ is formed by an OLS regression on the augmented design matrix from Section \ref{sec::mpt}. The matrix $A = \MosaicResid(Z, \bX) \in \R^{N \times T}$ is formed the same way. Therefore, Lemma \ref{lem::lm_fact} implies that for each cluster $C_m$,
\begin{equation}
    \hat\epsilon_{C_m}(b) = \hat\epsilon_{C_m}(0) - b A_{C_m}.
\end{equation}
Since this holds for every $m$, it implies that $\hat\epsilon(b) = \hat\epsilon(0) - b A$. 

Next, recall that $\tilde \epsilon(b), \tilde A$ are defined as $\tilde\epsilon_{C_m} = \hat\epsilon_{C_m} P^{B_m}, \tilde A_{C_m} = A_{C_m} P^{B_m}$ for $B_1, \dots, B_M \iid \Bern(0.5)$. As a result, we have for each $m \in [M]$,
\begin{equation}
    \tilde\epsilon_{C_m}(b) \defeq \hat\epsilon_{C_m}(b) P^{B_m} = \hat\epsilon_{C_m}(0) P^{B_m} - b A_{C_m} P^{B_m} = \tilde\epsilon_{C_m}(0) - b \tilde A_{C_m}.
\end{equation}
Since this holds for every $m$, we have that $\tilde\epsilon(b) = \tilde\epsilon(0) - b \tilde A$.


These results, taken together, show that
\begin{equation}
    S(\hat\epsilon(b)) \defeq \frac{\langle \hat\epsilon(b), D \rangle }{\langle D, D \rangle} = \frac{\langle \hat\epsilon(0), D \rangle}{\langle D, D\rangle} - b \frac{\langle A, D \rangle}{\langle D, D \rangle} = \hat\beta\mosaic - b,
\end{equation}
where the last equality uses $\langle A, D \rangle = \langle D, D \rangle$ (Lemma \ref{lem::angles}). Similarly,
\begin{equation}
    S(\tilde\epsilon(b)) \defeq \frac{\langle \tilde\epsilon(b), D \rangle}{\langle D, D \rangle} = \frac{\langle \tilde\epsilon(0), D \rangle}{\langle D, D\rangle} - b \frac{\langle \tilde A, D \rangle}{\langle D, D \rangle} = \tilde\beta - b \cdot \slope, 
\end{equation}
where the last equality uses $\langle \tilde A, D \rangle = \langle \tilde D, D \rangle$ (Lemma \ref{lem::angles}) and thus $\frac{\langle \tilde A, D \rangle}{\langle D, D \rangle} = \slope$.

As a result, we have that
\begin{align*}
    S(\hat\epsilon(b)) > S(\tilde\epsilon(b)) \Leftrightarrow \hat\beta\mosaic - b > \tilde\beta - b \cdot \slope \Leftrightarrow b < \frac{\hat\beta\mosaic - \tilde\beta}{1 - \slope} \Leftrightarrow b < \hat\beta\mosaic + \frac{\slope \hat\beta\mosaic - \tilde\beta}{1 - \slope}. 
\end{align*}
Therefore,
\begin{align}
    \P(\beta\opt \not \in \CI\mosaic) 
    &=
    \P\left(
    \left\{\beta\opt > \hat\beta\mosaic + \tilde{Q}_{1-\alpha/2}\left(\frac{\slope \hat\beta\mosaic - \tilde\beta}{1 - \slope} \right)\right\} \cup 
    \left\{\beta\opt < \hat\beta\mosaic + \tilde Q_{1-\alpha/2}\left(\frac{\slope \hat\beta\mosaic - \tilde\beta}{1 - \slope}\right)\right\}
    \right) \nonumber \\
    &\le 
        \P\left(S(\hat\epsilon(\beta\opt)) > \tilde Q_{1-\alpha/2}(S(\tilde\epsilon(\beta\opt))) \right) 
        + \P\left(S(\hat\epsilon(\beta\opt)) < \tilde Q_{\alpha/2}(S(\tilde\epsilon(\beta\opt))) \right) \label{eq::ci_equivalence} \\
    &\le 
        \alpha/2 + \alpha/2 = \alpha, \nonumber
\end{align}
where the last line follows because $\hat\epsilon(\beta\opt)$ satisfies joint invariance (see the proof of Theorem \ref{thm::test_finite_sample}) for the true value of $\beta\opt$, and thus the mosaic permutation test controls the Type I error rate.

\begin{remark}\label{rem::unbiased_bs_terms} Note that when the true errors are mean-zero, we have that
\begin{equation*}
    \E[\hat\beta\mosaic] = \E\left[\frac{\langle D, \hat\epsilon \rangle}{\langle D, D \rangle}\right] = \E\left[\frac{\langle D, \hat\epsilon(\beta\opt) - \beta\opt A \rangle}{\langle D, D \rangle}\right] = \beta\opt 
\end{equation*}
where again we use the fact that $\langle D, A \rangle = \langle A, A \rangle$ (Lemma \ref{lem::angles}). Furthermore,
\begin{equation*}
    \E[\tilde\beta \mid B_1,\dots, B_M] = \E\left[\frac{\langle D, \tilde\epsilon \rangle}{\langle D, D \rangle} \mid B_1, \dots, B_M\right] = \E\left[\frac{\langle D, \tilde\epsilon(\beta\opt) - \beta\opt \tilde A \rangle}{\langle D, D \rangle} \mid B_1, \dots, B_M\right] = \slope \beta\opt 
\end{equation*}
where again we use the fact that $\langle \tilde A, D \rangle = \langle \tilde D, D \rangle$ in the last step (Lemma \ref{lem::angles}). Since $\hat\beta\mosaic \Perp B_1, \dots, B_M$ and $\tilde\rho$ is a deterministic constant given $B_1, \dots, B_M$, this proves that
\begin{equation*}
\E\left[\frac{\slope \hat\beta\mosaic - \tilde\beta}{1-\slope} \mid B_1, \dots, B_M\right] = 0.
\end{equation*}
\end{remark}

We now prove the algebraic lemma, Lemma \ref{lem::angles}, used at the end of the proof of Theorem \ref{thm::ci_validity}.

\begin{lemma}\label{lem::angles} Let $A = \MosaicResid(Z, \bX)$, $D = \frac{1}{2} (A - AP)$, and $\tilde A = \MosaicRandomize(A), \tilde D = \MosaicRandomize(D)$. Then
\begin{equation}
    \langle A, D \rangle = \langle D, D \rangle \text{ and } \langle \tilde A, D \rangle = \langle \tilde D, D \rangle.
\end{equation}
\begin{proof} To prove the first result, recall $P$ is symmetric and $P^2 = I_T$. Thus we note the two equalities below:
\begin{equation*}
    \langle A, D \rangle = \trace(A D\trans) \defeq \trace\left(A \frac{1}{2} (I_T - P) A\trans\right).
\end{equation*}
\begin{equation*}
    \langle D, D \rangle = \trace(D D\trans) \defeq \trace\left(A \frac{1}{4} (I_T - P)^2 A\trans\right).
\end{equation*}
However, $(I_T - P)^2 = 2 (I_T - P)$, so $\frac{1}{4} (I_T - P)^2 = \frac{1}{2} (I_T - P)$ and thus $\langle A, D \rangle = \langle D, D \rangle.$ This proves the first result.

The second result is essentially similar. We note that by definition of $\tilde A$ and $\tilde D$,
\begin{equation*}
    \langle \tilde A, D \rangle = \sum_m \langle A_{C_m} P^{B_m}, D_{C_m} \rangle  \text{ and } 
    \langle \tilde D, D \rangle = \sum_m \langle D_{C_m} P^{B_m}, D_{C_m} \rangle.
\end{equation*}
Therefore it suffices to show that for any $m$, $\langle A_{C_m} P^{B_m}, D_{C_m} \rangle = \langle D_{C_m} P^{B_m}, D_{C_m} \rangle$ holds with probability one. The logic for the first result already proves this for the case where $B_m = 0$. When $B_m = 1$, we have
\begin{align*}
    \langle A_{C_m} P, D_{C_m} \rangle 
    = \trace\left(A_{C_m} P \frac{1}{2} (I_{|C_m|} - P) A_{C_m}\trans\right) = - \frac{1}{2} \trace(A_{C_m} (I_{|C_m|} - P) A_{C_m}\trans).
\end{align*}
\begin{align*}
    \langle D_{C_m} P, D_{C_m} \rangle = \frac{1}{4} \trace\left(A_{C_m} (I_{|C_m|} - P) P (I_{|C_m|} - P) A_{C_m}\trans \right) = - \frac{1}{2} \trace(A_{C_m} (I_{|C_m|} - P) A_{C_m} \trans) 
\end{align*}
where the last equality follows from the fact that $(I_{|C_m|} - P) P (I_{|C_m|} - P) = -2 (I_{|C_m|} - P).$ This proves $\langle A_{C_m} P, D_{C_m} = \langle D_{C_m} P, D_{C_m} \rangle$ which completes the proof.
\end{proof}
\end{lemma}

\section{Main proof details for robustness results}

\subsection{Proof of Proposition \ref{prop::finitesamplemoment}}\label{appendix::finitesamplemoment}

The proof proceeds in four steps. The first step performs basic calculations to simplify the definitions of $\Delta$ and $\tilde\Delta$. The second step simplifies $\mu_K$ and $\tilde\mu_K$. The third step shows $\E[\tilde\mu_K] \approx \E[\mu_K]$ via a combinatorial argument and Holder's inequality. The fourth step shows $\var(\tilde\mu_K) \to 0$ using an Efron-Stein argument.

Throughout the proof, we abbreviate $S \defeq S(\hat\epsilon)$ and $\tilde{S} \defeq S(\tilde\epsilon)$. We allow the value of $a_K$ to change from line to line, but in all cases its value depends only on $K$ and the sub-Gaussian norm of $\epsilon$.

\underline{Step 1: Simplifying $\Delta$ and $\tilde\Delta$.} 
Recall $S$ is defined as
\begin{equation*}
    S = 2 \sum_{m < m' \in [M]} \sum_{i \in C_m} \sum_{j \in C_{m'}} s_{ij} \hat\epsilon_i\trans \hat\epsilon_j.
\end{equation*}
$\tilde\epsilon$ replaces $\hat\epsilon_{C_m}$ with $\hat\epsilon_{C_m} P^{B_m}$ for each $m \in [M]$, so
\begin{align}
    \tilde S 
    &= 2 \sum_{m < m' \in [M]} \sum_{i \in C_m} \sum_{j \in C_{m'}} s_{ij} \hat\epsilon_i\trans P^{B_m + B_{m'}} \hat\epsilon_j \nonumber \\
    &= 
    2 \sum_{m < m' \in [M]} \sum_{i \in C_m} \sum_{j \in C_{m'}} \I(B_m = B_{m'}) s_{ij} \hat\epsilon_i\trans \hat\epsilon_j + \I(B_m \ne B_{m'}) s_{ij} \hat\epsilon\trans P \hat\epsilon_j. \label{eq::sopt_def}
\end{align}
where the first line uses the fact that $P$ is symmetric and the second uses the fact that, since $P^2 = I_T$, $P^{B_m + B_{m'}} = P^{\I(B_m \ne B_{m'})}$. Therefore, since $B_m \iid \Bern(1/2)$,
\begin{equation}
    \tilde{\E}[\tilde S] = \sum_{m' < m \in [M]} \sum_{i \in C_m} \sum_{j \in C_m} s_{ij} \hat\epsilon_i\trans \hat\epsilon_j + s_{ij} \hat\epsilon_i\trans P \hat\epsilon_j.
\end{equation}
Thus, if we define $Z_m \defeq 2 B_m - 1$ and $\delta_{m,m'} = \sum_{i \in C_m} \sum_{j \in C_{m'}} s_{ij} \hat\epsilon_i\trans (I_T - P) \hat\epsilon_j$, we conclude
\begin{equation}\label{eq::deltaM_comp}
    \Delta = \frac{S - \tilde{\E}[\tilde{S}]}{\sigma} = \frac{1}{\sigma} \sum_{m < m' \in [M]} \delta_{m,m'},
\end{equation}
\begin{equation}\label{eq::deltaMopt_comp}
    \tilde\Delta = \frac{\tilde{S} - \tilde{\E}[\tilde{S}]}{\sigma} = \frac{1}{\sigma} \sum_{m < m' \in [M]} Z_m Z_{m'} \delta_{m,m'}
\end{equation}
where $\sigma^2 \defeq \var(S - \tilde{\E}[\tilde{S}])$. We end by simplifying the expression for $\sigma$. In particular, note that $\delta_{m,m'}$ is a bilinear function of $\hat\epsilon_{C_m}, \hat\epsilon_{C_m'}$, and $\{\hat\epsilon_{C_m}\}_{m=1}^M$ are mean zero and jointly independent. Therefore, for any $m_0 < m_0' \in [M], m_1 < m_1' \in [M]$, we know $\E[\delta_{m_0,m_0'} \delta_{m_1, m_1'}] = 0$ unless $m_0 = m_1, m_0' = m_1'$. This implies
\begin{align}
        \sigma^2 
    &=
        \sum_{m_0 < m_0' \in [M]} \sum_{m_1 < m_1' \in [M]} \E\left[\delta_{m_0, m_0'} \delta_{m_1, m_1'} \right] 
    =
        \sum_{m < m' \in [M]} \E\left[\delta_{m, m'}^2\right].\label{eq::sigmaM_comp}
\end{align}

\underline{Step 2: Simplifying $\mu_K$ and $\tilde\mu_K$}. Let $\mcM = \{(m, m') : m < m' \in [M]\}$ denote the set of unique pairs of clusters, and $\mcM^K = \{g, h \in \N^K : (g_k, h_k) \in \mcM \text{ for } k=1, \dots, K\}$ denote its $K$th cartesian power.  By Eqs. \eqref{eq::deltaM_comp}-\eqref{eq::deltaMopt_comp}, we have that
\begin{equation}\label{eq::muK_expansion}
    \mu_K \defeq \E[\Delta^K] = \frac{1}{\sigma^K} \sum_{(g, h) \in \mcM^K} \E\left[\prod_{k=1}^K \delta_{g_k, h_k}\right].
\end{equation}
\begin{equation}\label{eq::muKopt_expansion}
    \tilde\mu_K \defeq \tilde{\E}[{\tilde\Delta}^K] = \frac{1}{\sigma^K} \sum_{(g, h) \in \mcM^K} \tilde{\E}\left[\prod_{k=1}^K Z_{g_k} Z_{h_k} \delta_{g_k, h_k}\right].
\end{equation}
To analyze these quantities, we must define two categories of index vectors $(g,h) \in \mcM^K$. In particular, for any cluster $m \in [M]$ and index vectors $g, h \in \mcM^K$, we let $\degree_m(g,h) \defeq \sum_{k=1}^K \I(g_k = m) + \I(h_k = m)$ denote the number of times that cluster $m$ appears in the index vectors $g, h$. 

We define the index vectors $(g,h)$ to be \textit{singletons} if there exists some cluster $m$ which has degree 1:
\begin{equation}\label{eq::singleton_def}
    \singleton \defeq \{(g,h) \in \mcM^K : \degree_m(g,h) = 1 \text{ for some } m \in [M]\}
\end{equation}
where $\nonsingleton \defeq \mcM^K \setminus \singleton$ is the set of non-singletons. We define $(g,h)$ to be \textit{even} if all degrees are even:
\begin{equation}\label{eq::even_def}
    \mcE = \{(g,h) \in \mcM^K : \degree_m(g,h) \text{ is even for all } m \in [M]\} \subset \nonsingleton.
\end{equation}
\underline{Step 2(a): analysis of $\mu_K$.} We now show that all terms in $\mu_K$ equal zero unless $(g,h) \in \nonsingleton$. To see this, fix any $(g,h) \in \mcS$. Then there exists some cluster $m$ such that $\degree_m(g,h) = 1$. Without loss of generality, assume $\degree_{g_1}(g,h) = 1$, so $g_2, \dots, g_K, h_1, \dots, h_K \ne g_1$. Then
\begin{equation}
    \E\left[\prod_{k=1}^K \delta_{g_k, h_k}\right] = \E\left[\left(\sum_{i \in C_{g_1}, j \in C_{h_1}} s_{ij} \hat\epsilon_i\trans (I_T - P) \hat\epsilon_j \right) \prod_{k=2}^K \delta_{g_k, h_k}\right] = 0
\end{equation}
where the final equality holds because $\{\hat\epsilon_i\}_{i \in C_{g_1}}$ is mean-zero and independent of all other quantities (since the quantities depend only data outside of cluster $g_1$ by assumption). Combining with Eq. \eqref{eq::muK_expansion} proves that
\begin{equation}\label{eq::muK_step1a}
    \mu_K = \frac{1}{\sigma^K} \sum_{(g,h) \in \nonsingleton} \E\left[\prod_{k=1}^K \delta_{g_k, h_k} \right].
\end{equation}
\underline{Step 2(b): analysis of $\tilde\mu_K$.} We now show that all terms in $\tilde\mu_K$ equal zero unless $(g,h) \in \mcE \subset \nonsingleton$. Since $\tilde{\E}$ takes the expectation over $Z_1, \dots, Z_M$ conditional on the data, we obtain
$$\tilde{\E}\left[\prod_{k=1}^K Z_{g_k} Z_{h_k} \delta_{g_k, h_k}\right] 
= \prod_{k=1}^K \delta_{g_k, h_k} \tilde{\E}\left[\prod_{m=1}^M (Z_m)^{\degree_m(g,h)}\right] = \begin{cases} \prod_{k=1}^K \delta_{g_k, h_k} & (g,h) \in \mcE \\ 
0 & (g,h) \not \in \mcE \end{cases}$$
where the last equality follows because $Z_1, \dots, Z_M \iid \Unif(\{-1,1\})$, and thus $\prod_{m=1}^M Z_m^{\degree_m(g,h)}$ is mean zero unless $\degree_m(g,h)$ is even for all $m$. When $\degree_m(g,h)$ is even for all $m$ (and thus $(g,h) \in \mcE$), we have that $\prod_{m=1}^M Z_m^{\degree_m(g,h)} = 1$ deterministically. Combining this result with Eq. \eqref{eq::muKopt_expansion} yields
\begin{equation}\label{eq::muKopt_step1b}
    \tilde\mu_K = \frac{1}{\sigma^K} \sum_{(g,h) \in \mcE} \prod_{k=1}^K \delta_{g_k, h_k}.
\end{equation}

\underline{Step 3: analysis of bias.} By Eqs. \eqref{eq::muK_step1a}-\eqref{eq::muKopt_step1b}, we obtain
\begin{equation}
    \underbrace{\mu_K - \E[\tilde\mu_K]}_{\text{bias}} = \frac{1}{\sigma^K} \sum_{(g,h) \in \nonsingleton \setminus \mcE} \E\left[\prod_{k=1}^K \delta_{g_k, h_k}\right].
\end{equation}
When $K = 2$, the bias is exactly zero, since in that case it turns out that $\nonsingleton = \mcE$, i.e., $(g,h) \in \mcM^2$ is even if and only if all degrees are zero or two. For larger $K$, the bias is nonzero, but we can bound the bias via Holder's inequality and a combinatorial argument.

In particular, by Holder's inequality and Jensen's inequality,
\begin{align}
        \left|\E\left[\prod_{k=1}^K \delta_{g_k, h_k} \right]\right|
    &\le 
        \prod_{k=1}^K \E\left[|\delta_{g_k, h_k}|^{K} \right]^{1/K} 
    \le 
        \prod_{k=1}^K \E\left[\delta_{g_k, h_k}^{2K} \right]^{1/2K} 
    \le 
        \max_{m < m' \in [M]} \sqrt{\E\left[\delta_{g_k, h_k}^{2K}\right]} \label{eq::max_norm_moment_bound}.
\end{align}
However, Lemma \ref{lem::delta_subexp} shows that since $\epsilon$ is sub-Gaussian, for any $\ell$, $\E[ \delta_{m, m'}^{\ell}] \le a_{\ell} T^{\ell/2}$ is uniformly bounded where $a_\ell$ depends only on $\ell$ and the sub-Gaussian constant. This implies that for some constant $a_K$ depending only on $K$ and the sub-Gaussian constant, 
\begin{equation}
    \left|\mu_K - \E[\tilde\mu_K]\right| \le \frac{a_K T^{K/2}}{\sigma^K} |\nonsingleton \setminus \mcE|.
\end{equation}
Next, a purely combinatorial argument in Lemma \ref{lem::odd_nonsingleton_combo} shows that $|\nonsingleton \setminus \mcE| \le c_K M^{K-1}$ for some constant $c_K$ depending only on $K$. Therefore
\begin{equation}\label{eq::bias_penultimate}
    \left|\mu_K - \E[\tilde\mu_K]\right| \le \frac{a_K T^{K/2}}{\sigma^K} M^{K-1}.
\end{equation}
However, combining Eq. \eqref{eq::sigmaM_comp} and the definition of $\sigma_{\delta}^2$ from Eq. \eqref{eq::sigma_delta_def} yields
\begin{equation}\label{eq::sigma2_bound_useful}
    \sigma^2 = \sum_{m < m' \in [M]} \E[\delta_{m, m'}^2] = \sum_{m < m' \in [M]} \var(\delta_{m, m'}) = T \cdot \binom{M}{2} \cdot \sigma_{\delta}^2 \ge T \frac{M^2}{4} \sigma_{\delta}^2.
\end{equation}
Thus, we conclude that $\sigma^K \ge 2^{-K} M^K T^{K/2} \sigma_{\delta}^K$. Plugging this into Eq. \eqref{eq::bias_penultimate} yields the bias result:
\begin{equation}\label{eq::bias_ultimate}
    \left|\mu_K - \E[\tilde\mu_K]\right| \le \frac{a_K}{\sigma_{\delta}^K M},
\end{equation}

\underline{Step 4: analysis of variance.} $\tilde\mu_K$ is a function of $\{\epsilon_{C_m}\}_{m=1}^M$ which are jointly independent. Let $\{\epsilon_{C_m}\opt\}_{m=1}^M$ denote independent draws from the law of $\{\epsilon_{C_m}\}_{m=1}^M$ and let $\tildemuKm$ denote the value of $\tilde\mu_K$ after replacing $\epsilon_{C_m}$ with $\epsilon_{C_m}\opt$. The Efron-Stein inequality yields that
\begin{equation}
    \var(\tilde\mu_K) \le \frac{1}{2} \sum_{m=1}^M \E\left[\left(\tilde\mu_K - \tildemuKm\right)^2\right].
\end{equation}
For convenience, let $\Pi(g,h) = \prod_{k=1}^K \delta_{g_k, h_k}$ and let $\Pi^{(m)}(g,h)$ be defined similarly but replacing $\epsilon_{C_m}$ with $\epsilon_{C_m}\opt$. Note that $\Pi(g,h) = \Pi\upm(g,h)$ whenever $\degree_m(g,h) = 0$, since in this case cluster $m$ is not involved in $\Pi(g,h)$. Let $\mcE_m = \{(g,h) \in \mcE : \degree_m(g,h) > 0\}$ denote the set of index vectors such that $\Pi(g,h) \ne \Pi\upm(g,h)$. Then
\begin{equation}
    \tilde\mu_K - \tildemuKm = \sigma^{-K} \sum_{(g,h) \in \mcE_m} \Pi(g,h) - \Pi\upm(g,h),
\end{equation}
because Eq. \eqref{eq::deltaMopt_comp} proves $\tilde\mu_K = \sigma^{-K} \sum_{(g,h) \in \mcE} \Pi(g,h)$ and $\tildemuKm = \sigma^{-K} \sum_{(g,h) \in \mcE} \Pi^{(m)}(g,h)$. Thus
\begin{equation}\label{eq::efron_stein_simplified}
    \var(\tilde\mu_K) \le \frac{1}{2 \sigma^{2K}} \sum_{m=1}^M \sum_{(g,h) \in \mcE_m} \sum_{(g', h') \in \mcE_m} \E\left[(\Pi(g,h) - \Pi^{(m)}(g,h)) (\Pi(g',h') - \Pi^{(m)}(g',h'))\right].
\end{equation}

Now, we apply moment bounds and a combinatorial bound on $|\mcE_m|$ to bound $\var(\tilde\mu_K)$. To start, note that for any $(g,h), (g', h') \in \mcE_m$, Holder's inequality and the $C_r$-inequality yield
\begin{align*}
        \E\left[(\Pi(g,h) - \Pi^{(m)}(g,h)) (\Pi(g',h') - \Pi^{(m)}(g',h'))\right] 
    &\le
        \sqrt{\E\left[(\Pi(g,h) - \Pi^{(m)}(g,h))^2\right] \E\left[(\Pi(g',h') - \Pi^{(m)}(g',h'))^2\right]} \\
    &\le 
        4 \sqrt{\E\left[(\Pi(g,h))^2\right] \E\left[(\Pi(g',h'))^2\right]}
\end{align*}
where above, we also use the fact that $\Pi(g,h) \sim \Pi\upm(g,h)$ have the same marginal distribution. Furthermore, as argued in Eq. \eqref{eq::max_norm_moment_bound}, $\E[\Pi(g,h)^2] \le \max_{m < m' \in [M]} \sqrt{\E[\delta_{m,m'}^{4K}]} \le T^K a_K$ for some constant $a_K$ depending only on $K$ and the sub-Gaussian constant (see also Lemma \ref{lem::delta_subexp}). Combining with Eq. \eqref{eq::efron_stein_simplified} yields
\begin{equation}
    \var(\tilde\mu_K) \le \frac{a_K T^K}{\sigma^{2K}} \sum_{m=1}^M |\mcE_m|^2.
\end{equation}
However, a combinatorial argument in Lemma \ref{lem::efronstein_combo} shows that there exists a universal constant $c_K$ depending only on $K$ such that $|\mcE_m| \le c_K M^{K-1}$. Therefore
\begin{equation}
    \var(\tilde\mu_K) \le \frac{a_K T^K}{\sigma^{2K}} M^{2K-1}.
\end{equation}
In Step 3, we argued that $\sigma \ge \frac{M}{2} \sqrt{T} \sigma_{\delta}$. Therefore we conclude
\begin{equation}
    \var(\tilde\mu_K) \le \frac{a_K}{\sigma_{\delta}^{2K} M}.
\end{equation}
Combining this with Eq. \eqref{eq::bias_ultimate} proves that
\begin{equation}
    \E[(\tilde\mu_K - \mu_K)^2] = \var(\tilde\mu_K) + \E[(\mu_K - \tilde\mu_K)^2] \le \frac{a_K}{\sigma_{\delta}^{2K} M}
\end{equation}
where $a_K$ depends only on $K$ and the sub-Gaussian constant $\sgconstant$.


\begin{remark}[Sub-gaussianity]\label{rem::subg} The sub-Gaussianity assumption is only used to show that $\E[T^{-K/2} \delta_{m,m'}^K]$ exists and is uniformly bounded in the number of clusters $M$. Instead, we may directly assume that $\E[\delta_{m,m'}^K] \le T^{K/2} B_K$ is uniformly bounded in $M$ for each $K$. Since $B_k$ may grow arbitrarily with $k$, this assumption actually allows $\delta_{m,m'}$ to be arbitrarily heavy-tailed, as long as (a) all of its moments exist and (b) it does not become progressively \textit{more} heavy-tailed as $M \to \infty$. Under this assumption, Proposition \ref{prop::finitesamplemoment} holds without any modifications.
\end{remark}

\subsection{Proof of Theorem \ref{thm::asymptotic_result}}


We now state a general result showing that under regularity conditions, if the moments of $F_M$ converge to those of $\tilde F_M$, then $\sup_{x \in \R}|F_M(x) - \tilde F_M(x)| \toprob 0$ holds. Then, Type I error control follows from Lemma A.1 of \cite{romano2012}.

\begin{proposition}\label{prop::strategy} Let $F_M$ be a sequence of CDFs and let $\tilde F_M$ denote a sequence of random CDFs defined on a common probability space. Let $\mu_K(F_M), \mu_K(\tilde F_M)$ denote their $K$th moments, and assume the following:
\begin{enumerate}[(a)]
    \item Bounded moments: for each $K \in \N$, $\limsup_{M \to \infty} \mu_K(F_M) \le L^K K!$ for some $L \ge 0$.
    \item Convergence of moments: for each $k \in \N$, $\mu_k(\tilde F_M) - \mu_k(F_M) \toprob 0$ as $M \to \infty$.
    \item Anticoncentration: for some $\xi > 0$, $\limsup_{M \to \infty} |F_M(b) - F_M(a)| \le \xi |a-b|$ for all $a, b \in \R$.
\end{enumerate}
Then we have three results. First,
\begin{equation}
    \KS_M \defeq \sup_{x \in \R} |F_M(x) - \tilde F_M(x)| \toprob 0.
\end{equation}
Second, if $\Delta_M \sim F_M$, we have that
\begin{equation}
    \limsup_{M \to \infty} \P(\Delta_M > {\tilde F_M}^{-1}(1-\alpha)) \le \alpha.
\end{equation}
Third, if additionally we assume $\limsup_{M \to \infty} \P(\Delta_M = {\tilde F_M}^{-1}(1-\alpha)) = 0$, then 
\begin{equation}
    \lim_{M \to \infty} \P(\Delta_M > {\tilde F_M}^{-1}(1-\alpha)) = \alpha.
\end{equation}
\begin{proof} The proof is in three steps.

\underline{Step 1: Subsequence tricks.} Recall that a sequence of random variables $\{\KS_M\}_{M=1}^{\infty}$ converges in probability to zero if and only if for every subsequence $M_n$, there exists a further subsequence $\{M_{n_j}\}$ such that $\KS_{M_{n_j}} \toas 0$. To show the result, fix any subsequence $M_n$. We note the following:
\begin{itemize}[topsep=0pt]
    \item By Assumption (a), the sequence $\{F_{M_n}\}$ has uniformly bounded moments and is tight. By Prokhorov's theorem, there exists a subsubsequence $F_{M_{n_j}} \tod F$ which congerges in law to $F$. 
    \item By Assumption (b), for each $K \in \N$,  there exists a subsubsequence $\{M_{n_j}\}$ such that $\mu_K(\tilde F_{M_{n_j}}) - \mu_K(F_{M_{n_j}}) \toas 0$.
\end{itemize}
By a standard diagonalization argument (as used in, e.g., Helly's selection theorem), we may pick a common subsequence $M_{n_j}$ along which all of the above convergences occur. Formally, this means that (i) $F_{M_{n_j}} \tod F$ holds deterministically and (ii) with probability one, the limits $\lim \mu_K(F_{M_{n_j}}) - \mu_K(\tilde F_{M_{n_j}})$ exist and equal zero for all $K \in \N$ simultaneously.

\underline{Step 2: Applying Billingsley Theorem 30.2}. We now apply Billingsley Theorem 30.2 \citep{billingsley1995probability}, which tells us that if $F$ is determined by its moments and $\mu_K(\tilde F_{M_{n_j}}) \to \mu_K(F)$ for every $K \in \N$ simultaneously, then $\tilde F_{M_{n_j}} \tod F$. To establish the conditions of this theorem, we make the following notes:
\begin{enumerate}[(i)]
    \item \emph{The moments of $F_{M_{n_j}}$ converge.} Since $\mu_K(F_{M_{n_j}})$ is uniformly bounded by Assumption (a) for all $K$, the Vitali convergence theorem and the result from Step 1 that $F_{M_{n_j}} \tod F$ imply $\mu_K(F_{M_{n_j}}) \to \mu_K(F)$. 
    \item \emph{The moments of $\tilde F_{M_{n_j}}$ converge.} By observation (i) and the result from Step 1 that $\lim \mu_K(F_{M_{n_j}}) - \mu_K(\tilde F_{M_{n_j}}) = 0$ simultaneously for all $K \in \N$, this implies that $\mu_K(\tilde F_{M_{n_j}}) \to \mu_K(F)$ holds simultaneously for all $K \in \N$ with probability one. 
    \item \emph{$F$ is determined by its moments}: Combining observation (i) with Assumption (a), we conclude $\mu_K(F) = \lim \mu_K(F_{M_{n_j}}) \le L^K K!$. Thus, $F$ has a moment generating function with a positive radius of convergence; thus, $F$ is determined by its moments.
\end{enumerate}

Thus, we have established that the assumptions of Billingsley 30.2 hold with probability one. We conclude that with probability one, $\tilde F_{M_{n_j}} \tod F$. I.e., $\P(\tilde F_{M_{n_j}}(x) \to F(x) \text{ for all } x \in \R) = 1$.

\underline{Step 3: Applying Polya's theorem.} Note that $|F(b) - F(a)| = \lim |F_{M_{n_j}}(b) - F_{M_{nj}}(a)| \le \xi |a-b|$ for any $a, b \in \R$ that are continuity points of $F$, by Assumption (c) and the fact that $F_{M_{n_j}} \tod F$. Yet $F$ is a CDF, so it is continuous almost everywhere. Therefore this result implies that $F$ cannot have point masses, and thus it is a continuous CDF. By Polya's theorem and the result of Step 2, we conclude that
\begin{equation}
    \KS_{M_{n_j}} = \sup_{x \in \R} |F_{M_{n_j}}(x) - \tilde F_{M_{n_j}}(x)| \toas 0.
\end{equation}
This completes the proof of the first result. The next two results follow essentially immediately from Lemma A.1 of \cite{romano2012}. In particular, fix any $\epsilon > 0$ and note that for any $\delta > 0$, for sufficiently large $M$, we have that $\P(\KS_M \ge \epsilon) \ge 1 - \delta$. 
\cite{romano2012}, Lemma A.1 result (vi) proves that under this condition, we have that
\begin{equation*}
    \P(\Delta_M \le {\tilde F_M}^{-1}(1-\alpha)) \ge 1 - \alpha - \epsilon - \delta. 
\end{equation*}
By taking $\epsilon, \delta$ sufficiently small, we find that
\begin{equation*}
    \limsup_{M \to \infty} \P(\Delta_M > {\tilde F_M}^{-1}(1-\alpha)) = 1 - \liminf_{M \to \infty} \P(\Delta_M \le {\tilde F_M}^{-1}(1-\alpha)) \le \alpha.
\end{equation*}
For the final result, \cite{romano2012}, Lemma A.1 result (vii) proves that if $\P(\KS_M \ge \epsilon) \ge 1 - \delta$,
\begin{equation*}
    \P(\Delta_M \ge {\tilde F_M}^{-1}(1-\alpha)) \ge \alpha - \epsilon - \delta.
\end{equation*}
If $\lim_{M \to \infty} \P(\Delta_M = {\tilde F_M}^{-1}(1-\alpha)) = 0$, we conclude
\begin{equation*}
    \liminf_{M \to \infty} \P(\Delta_M > {\tilde F_M}^{-1}(1-\alpha)) = \liminf_{M \to \infty} \P(\Delta_M \ge {\tilde F_M}^{-1}(1-\alpha)) \ge \alpha.
\end{equation*}
\end{proof}
\end{proposition}

We now prove Theorem \ref{thm::asymptotic_result}, which is restated below for convenience.

\begingroup
\def\thetheorem{\ref{thm::asymptotic_result}}
\begin{theorem} Suppose Assumption \ref{assump::meanzero} (mean-zero errors), Assumption \ref{assump::subgaussian} (sub-Gaussianity) and the null $\mcH_0$ hold for every dataset $Y\upM$ in the triangular array. 
Then under Assumptions \ref{assump::minsigma}-\ref{assump::anticoncentration}, we have
\begin{equation}
    \limsup_{M \to \infty} \P(S_M > \tilde Q_{1-\alpha}(\tilde S_M)) \le \alpha.
\end{equation}
Furthermore, if $\lim_{M \to \infty} \P(S_M = \tilde Q_{1-\alpha}(\tilde S_M)) = 0$, then Eq. \eqref{eq::main_asymptotic_result} holds with equality.
\begin{proof} Since $\P(S_M > \tilde Q_{1-\alpha}(\tilde S_M)) = \P(\Delta_M > \tilde Q_{1-\alpha}(\tilde\Delta_M))$, it suffices to verify the conditions of Proposition \ref{prop::strategy}, where $F_M$ denotes the CDF of $\Delta_M$ and $\tilde F_M$ denotes the CDF of $\tilde\Delta_M$ conditional on the data. Note the following:
\begin{itemize}
    \item Assumptions \ref{assump::meanzero}-\ref{assump::subgaussian} and the null $\mcH_0$ hold for each $M$. Thus, we may apply Proposition \ref{prop::finitesamplemoment}, which tells us that for each $M$,
    \begin{equation}
\E\left[\left(\mu_K(F_M) - \mu_K(\tilde F_M) \right)^2\right] \le \frac{a_K}{\sigma_{\delta,M}^{2K} M}.
    \end{equation}
    Since $a_K$ does not depend on $M$ and $\liminf_{M \to \infty} \sigma_{\delta,M}^2 > 0$, $\mu_K(F_M) - \mu_K(\tilde F_M) \toprob 0$ for all $K$.
    \item Lemma \ref{lem::Delta_subexp} shows that $\Delta_M$ is sub-exponential, with sub-exponential norm depending only on $\sgconstant$ and not depending on $M$. This implies that $\mu_K(M) \le L^K K!$ for some $L$.
    \item We directly assume the anticoncentration assumption via Assumption \ref{assump::anticoncentration}.
\end{itemize}
Thus, we may apply Proposition \ref{prop::strategy}, which completes the proof.
\end{proof}
\end{theorem}
\endgroup

\subsection{Proof of Theorem \ref{thm::robust_cis}}\label{appendix::robustcis}

\textbf{Notation}: let $\hat\epsilon\upM(b) = \MosaicResid(Y\upM - b Z\upM, \bX\upM)$ denote the mosaic residuals for the $M$th element of the triangular array after subtracting $b Z\upM$ off of the outcomes $Y\upM$. $A = \MosaicResid(Z\upM, \bX\upM)$ are the values of the covariate of interest after regressing out the influence of $\bX\upM$ using a mosaic regression. Let $\CI\mosaic\upM$ be the mosaic confidence interval, and let $\beta\optM$ denote the true value of the coefficient for the $M$th element of the triangular array. We define $\tilde\epsilon\upM(b) = \MosaicRandomize(\hat\epsilon\upM(b))$ and $\tilde A\upM = \MosaicRandomize(A)$. 
To ease readability, we abuse notation and let $\hat\epsilon \defeq \hat\epsilon\upM(\beta\optM)$ and $\tilde\epsilon \defeq \tilde\epsilon\upM(\beta\optM)$ 
denote the value of the residuals and the transformed residuals, respectively, using the true value $b = \beta\optM$ of the coefficient of interest. Furthermore, let $S_M$ and $\tilde S_M$ denote the values of the test statistic applied to $\hat\epsilon$:
\begin{equation}
    S_M \defeq S(\hat\epsilon) \text{ and }  \tilde S_M \defeq S(\tilde\epsilon).
\end{equation}
Following Section \ref{appendix::finitesamplemoment}, we define $\Delta_M$ and $\tilde \Delta_M$ to be the appropriately normalized variants of $S_M$ and $\tilde S_M$:
\begin{equation}
    \Delta_M \defeq \frac{S_M - \tilde\E[\tilde S_M]}{\sqrt{\var(S_M - \tilde \E[S_M])}} \text{ and } \tilde\Delta_M \defeq \frac{\tilde S_M -\tilde\E[\tilde S_M]}{\sqrt{\var(S_M - \tilde \E[S_M])}}.
\end{equation}

\textbf{Proof strategy}: Note that Eq. \eqref{eq::ci_equivalence} proves that $\beta\optM \not \in \CI\mosaic\upM$ if and only if $S_M > \tilde Q_{1-\alpha/2}(\tilde S_M)$ or $S_M < \tilde Q_{1-\alpha/2}(\tilde S_M)$. Therefore, it suffices to show that for any $\alpha \in (0,1)$.
\begin{equation}
    \lim_{M \to \infty} \P(S_M < \tilde Q_{\alpha}(\tilde S_M)) \le \alpha \text{ and } \lim_{M \to \infty} \P(S_M > \tilde Q_{1-\alpha}(\tilde S_M)) \le \alpha.
\end{equation}
Note that $S_M < \tilde Q_{\alpha}(\tilde S_M)$ if and only if $\Delta_M < \tilde Q_{\alpha}(\tilde \Delta_M)$, so it also suffices to show the result above replacing $S_M$ and $\tilde S_M$ with $\Delta_M$ and $\tilde \Delta_M$, respectively. 

This proof is far easier than the proof of Theorem \ref{thm::asymptotic_result}, since in this case $\Delta_M \tod \mcN(0,1)$. Thus, the proof proceeds in three steps:
\begin{itemize}[itemsep=0.5pt, topsep=0pt, leftmargin=*]
    \item Step 1: Simplifying the definitions of $\Delta_M$ and $\tilde \Delta_M$.
    \item Step 2: Showing that $\Delta_M \tod \mcN(0,1)$.
    \item Step 3: Showing that if $\tilde F_M$ is the CDF of $\tilde \Delta_M$ conditional on the data, then $\sup_{x \in \R} |\tilde F_M(x) - \Phi(x)| \toas 0$.
    \item Step 4: Showing the final result via Lemma A.1 of \cite{romano2012}.
\end{itemize}
We now show these results in order. It may be helpful to recall that all elements of the triangular array are defined on the same probability space by assumption.

\textbf{Step 1: Simplifications.} Recall that if $D \defeq A - AP$, then
\begin{equation}
    S_M = \frac{\langle D, \hat\epsilon \rangle}{\langle D, D \rangle} = \frac{1}{\langle D, D \rangle} \sum_{m \in [M]} \langle D_{C_m}, \hat\epsilon_{C_m} \rangle \text{ and } \tilde S_M = \frac{\langle D, \tilde\epsilon \rangle}{\langle D, D \rangle} = \frac{1}{\langle D, D \rangle} \sum_{m \in [M]} \langle D_{C_m}, \hat\epsilon_{C_m} P^{B_m} \rangle.
\end{equation}
Since $B_1, \dots, B_M \iid \Bern(0.5)$, this implies that
\begin{equation}
    \tilde\E[\tilde S_M] = \frac{1}{2 \langle D, D \rangle} \sum_{m \in [M]} \langle D_{C_m}, \hat\epsilon_{C_m} (I_T + P) \rangle
\end{equation}
and therefore, up to nonrandom constants, we have that
\begin{equation}
    S_M  - \tilde\E[\tilde S_M] \propto \sum_{m \in [M]} \langle D_{C_m}, \hat\epsilon_{C_m} (I_T - P) \rangle = \sum_{m \in [M]} \hat\theta_M.
\end{equation}
\begin{equation}
    \tilde S_M - \tilde \E[\tilde S_M] \propto \sum_{m \in [M]} Z_m \hat\theta_M
\end{equation}
where above, $\hat \theta_m \defeq \langle D_{C_m}, \hat\epsilon_{C_m} (I_T - P) \rangle$ and $Z_m \defeq 2B_m - 1 \simiid \Unif(\{-1,1\})$. This implies that if we define $\sigma_M \defeq \sqrt{\var(\sum_{m \in [M]} \hat\theta_M)}$, then
\begin{equation}
    \Delta_M = \frac{\sum_{m \in [M]} \hat\theta_M}{\sigma_M} \text{ and } \tilde \Delta_M \defeq \frac{\sum_{m \in [M]} Z_M \hat\theta_M}{\sigma_M}.
\end{equation}

\textbf{Step 2: Asymptotic normality of $\Delta_M$.} $\hat\theta_m$ is a (linear) function of $\hat\epsilon_{C_m}$ which in turn is a deterministic (linear) function of $\epsilon_{C_m}$. Under cluster independence, $\{\epsilon_{C_m}\}_{m=1}^M$ are independent, and we assume that $\epsilon$ is mean zero. Taken together, this implies that $\{\hat\theta_m\}_{m \in [M]}$ are jointly independent and mean zero.

We directly assume that the Lyapunov condition holds, namely, that for some $\delta > 0$, $\frac{1}{\sigma_M^{2+\delta}} \sum_{m=1}^M \E[|\hat\theta_m|^{2+\delta}] \to 0$. As a result, the Lyapunov CLT implies that
\begin{equation}
    \Delta_M \tod \mcN(0,1).
\end{equation}

\textbf{Step 3: Conditional asymptotic normality of $\tilde \Delta_M$.} Let $\tilde F_M(x) \defeq \tilde P(\tilde \Delta_M \le x)$ denote the conditional law of $\tilde \Delta_M$ given the data. Let $G$ be the event that the sequence of (random) CDFs $\{\tilde F_M\}_{M=1}^{\infty}$ converges in law to $\Phi(x)$. We begin by showing that $\P(G) = 1$.

\underline{Step 3a}: As an intermediate step, define $\sigma_M\opt \defeq \sqrt{\sum_{m \in [M]} \hat\theta_m^2} = \sqrt{\tilde\E\left[\left(\sum_{m \in [M]} \hat\theta_m Z_m \right)^2\right]}$ and $\Delta_M\opt = \frac{\sum_{m \in [M]} Z_m \hat\theta_m}{\sigma_M\opt}$. Note that conditional on the data, $\{Z_m \hat\theta_m\}_{m=1}^M$ are mean-zero, independent random variables satisfying $\tilde\E[|\hat\theta_m Z_m|^{2+\delta}] = |\hat\theta_m|^{2+\delta}$. Furthermore, we have by Lemma \ref{lem::uniform_ratio_consistency} and the moment conditions in Assumption \ref{assump::lyapunov} that
\begin{equation}\label{eq::conditional_lyapunov}
    \frac{\sum_{m \in [M]} |\hat\theta_m|^{2+\delta}}{{\sigma_M\opt}^{2+\delta}} \toas \lim_{M \to \infty} \frac{\E[|\hat\theta_m|^{2+\delta}]}{\sigma_M^{2+\delta}} = 0.
\end{equation}
As a result, the Lyapunov condition for the sequence $\{\Delta_M\opt\}_{M=1}^{\infty}$ holds asymptotically with probability one. Therefore, if $F_M\opt(x) \defeq \tilde P(\Delta_M\opt \le x)$ is the conditional CDF of $\Delta_M\opt$ given the data and $G\opt$ denotes the event that $\{F_M\opt\}_{M=1}^{\infty}$ converges weakly to $\mcN(0,1)$, we conclude that $\P(G\opt) = 1$ (which follows since $G\opt$ is implied by the a.s. convergence in Eq. \ref{eq::conditional_lyapunov}).

\underline{Step 3b}: Let $G_\sigma$ denote the event that $\lim_{M \to \infty} \frac{\sigma_M}{\sigma_M\opt} = 1$. We will now show that the events $G_{\sigma}$ and $G\opt$ together imply the event $G$. This is because $\tilde \Delta_M = \frac{\sigma_M\opt}{\sigma_M} \Delta_M\opt$, so therefore
\begin{equation}
    \tilde F_M(x) = \tilde \P(\tilde \Delta_M \le x) = \tilde\P\left(\frac{\sigma_M\opt}{\sigma_M} \Delta_M\opt \le x \right) = F_M\opt\left(x \cdot \frac{\sigma_M}{\sigma_M\opt} \right).
\end{equation}
On the event $G\opt$, we have that for all $x \in \R$,
\begin{equation}
    \lim_{M \to \infty} F_M\opt\left(x \cdot \frac{\sigma_M}{\sigma_M\opt} \right) = \lim_{M \to \infty}  \Phi\left(x \frac{\sigma_M}{\sigma_M\opt} \right). 
\end{equation}
However, on the event $G_{\sigma}$, we have that $\lim_{M \to \infty} \frac{\sigma_M}{\sigma_M\opt} = 1$. Since the normal CDF is everywhere continuous, this implies that on the event $G \cap G_{\sigma}$,
\begin{equation}
    \tilde F_M(x) = \lim_{M \to \infty} F_M\opt\left(x \cdot \frac{\sigma_M}{\sigma_M\opt} \right) = \lim_{M \to \infty}  \Phi\left(x \frac{\sigma_M}{\sigma_M\opt} \right) = \Phi(x)
\end{equation}
where the above holds simultaneously for all $x \in \R$. Therefore, by definition of weak convergence, this proves that $G\opt \cap G_{\sigma}$ implies $G$. Since $\P(G\opt) = 1$ and $\P(G_{\sigma}) = 1$ (since $\frac{\sigma_M}{\sigma_M\opt} \toas 1$ by Lemma \ref{lem::uniform_ratio_consistency} plus the continuous mapping theorem), this proves $\P(G) = 1$.

On the event $G$, $\{\tilde F_M\}_{M=1}^{\infty}$ converges weakly to $\mcN(0,1)$. By Polya's theorem, on the event $G$, we have that $\lim_{M \to \infty} \sup_{x \in \R} |\tilde F_M(x) - \Phi(x)| = 0$. Therefore, we conclude $\sup_{x \in \R} |\tilde F_M(x) - \Phi(x)| \toas 0$.

\textbf{Step 4: proving the final result}. Let $F_M$ be the CDF of $\Delta_M$ and note that
\begin{equation}   
    \sup_{x \in \R} |\tilde F_M(x) - F_M(x)| \le \sup_{x \in \R} |\tilde F_M(x) - \Phi(x)| + |\Phi(x) - F_M(x)| \toas 0.
\end{equation}
As a result, by Lemma A.1 of \cite{romano2012} (see, e.g., the end of proof of Theorem \ref{prop::strategy} for details), we conclude the desired result:
\begin{equation}
    \lim_{M \to \infty} \P(\Delta_M < \tilde Q_{\alpha}(\tilde \Delta_M)) \le \alpha \text{ and } \lim_{M \to \infty} \P(\Delta_M > \tilde Q_{1-\alpha}(\tilde \Delta_M)) \le \alpha.
\end{equation}

\section{Technical proofs}

\subsection{Moment bounds based on sub-Gaussianity}

Recall a random vector $\xi \in \R^n$ is sub-Gaussian if it has finite Luxemburg norm:
\begin{equation}
    \|\xi\|_{\psi_2} = \inf\{K > 0 : \sup_{t \in S_{n-1}} \E\left[\exp\left(\langle t, \xi \rangle^2 / K^2\right) \right] \le 2 \}
\end{equation}
where $\inf\emptyset = \infty$. Similarly, a random matrix $\xi \in \R^{n \times m}$ is sub-Gaussian if $\vecop(\xi)$ is sub-Gaussian. A random variable $\zeta \in \R$ is sub-exponential if the following norm is finite:
\begin{equation}
    \|\zeta\|_{\psi_1} = \inf\left\{K > 0 : \E\left[\exp\left(\left|\zeta/K\right|\right)\right] \le 2 \right\}.
\end{equation}
We now review a result from \cite{zajkowski2020}.

\begin{proposition}\label{prop::subg2subexp} For $\xi \in \R^n$ sub-Gaussian and any $A \in \R^{n \times n}$, then $\|\xi^T A \xi\|_{\psi_1} \le \|\xi\|_{\psi_2}^2 \trace(A A^T)^{1/2}.$
\end{proposition}
We now apply this result to show that both $\delta_{m,m'}$ and $\Delta$ are sub-exponential whenever $\epsilon$ is sub-Gaussian. First we show that the mosaic estimator $\hat\epsilon$ is sub-Gaussian.

\begin{lemma}\label{lem::linear_sub} For $\xi \in \R^n$ and $A \in \R^{k \times n}$, $\|A \xi\|_{\psi_2}\le \|A^T\|_{\mathrm{op}} \|\xi\|_{\psi_2}.$
\begin{proof} It suffices to show that if $K$ satisfies $\E[(\langle t, \xi\rangle)^2 / K^2] \le 2$ for all $t \in S^{n-1}$, then $\E[(\langle r, A\xi\rangle)^2 / (\|A\|_{\mathrm{op}} K)^2] \le 2$ also holds for all $r \in S^{k-1}$. However, for any $r \in S^{k-1}$, we have that
\begin{equation}
    \E\left[\left(\frac{\langle r, A \xi \rangle}{K \|A^T\|_{\mathrm{op}}}\right)^2\right] \le 
    \E\left[\left(\frac{\langle r, A \xi \rangle}{K \|A^T r \|_2}\right)^2\right]
        =
    \E\left[\left(\left\langle \frac{A^T r}{\|A^T r\|_2}, \xi \right\rangle \right)^2 / K^2 \right] \le 2
\end{equation}
where the first inequality follows because $\|A^T r\|_2 \le \|A\|_{\mathrm{op}} \|r\|_2 = \|A\|_{\mathrm{op}}$ and the second equality follows since $\|\xi\|_{\psi_1} = K$.
\end{proof}
\end{lemma}
\begin{corollary}\label{cor::hateps_subG} $\|\hat\epsilon\|_{\psi_2} \le \|\epsilon\|_{\psi_2}$.
\begin{proof} This follows from the previous lemma since $\vecop(\hat\epsilon) = H_{\mathrm{mosaic}} \vecop(\epsilon)$ where $H_{\mathrm{mosaic}}$ is the mosaic (block-diagonal) projection matrix.
\end{proof}
\end{corollary}

With this result, we now show that $\frac{1}{\sqrt{T}} \delta_{m,m'}$ is sub-exponential.

\begin{lemma}\label{lem::delta_subexp} Under Assumptions \ref{assump::meanzero}-\ref{assump::subgaussian}, for any $m \ne m' \in [M]$, $\left\|\frac{1}{\sqrt{T}} \delta_{m, m'}\right\|_{\psi_1} \le 2 \|\epsilon\|_{\psi_2}.$ Thus, there exists a constant $C$ depending only on $\|\epsilon\|_{\psi_2}$ such that
\begin{equation}
    T^{-K/2} \E[\delta_{m,m'}^K] \le K! C^K.
\end{equation}
\begin{proof} Observe that
\begin{equation*}
    \delta_{m,m'} = \sum_{i \in C_m} \sum_{j \in C_{m'}} s_{ij} \hat\epsilon_i^T (I_T - P)  \hat\epsilon_j = \sum_{t \in [T]} \sum_{t' \in [T]} \sum_{i \in C_m} \sum_{j \in C_{m'}} s_{ij} ([I_T]_{t, t'} - P_{t,t'}) \hat\epsilon_{i,t} \hat\epsilon_{j,t'}.
\end{equation*}
This is a quadratic form in $\hat\epsilon$, so we may apply Proposition \ref{prop::subg2subexp}. This yields the bound 
\begin{equation}
    \|\delta_{m,m'}\|_{\psi_1} \le \sqrt{\sum_{i \in C_m} \sum_{j \in C_m} s_{ij}^2 \|I_T - P\|_{\mathrm{Fr}}^2} \|\hat\epsilon\|_{\psi_2}^2.
\end{equation}
By Corollary \ref{cor::hateps_subG}, we know $\|\hat\epsilon\|_{\psi_2} \le \|\epsilon\|_{\psi_2}$. Furthermore, we know that $\|I_T - P\|_{\mathrm{Fr}} \le \|I_T\|_{\mathrm{Fr}} + \|P\|_{\mathrm{Fr}} = 2 \sqrt{T}$ by the triangle inequality and the fact that $P^2 = I_T$. Applying this yields
\begin{equation}
    \|\delta_{m,m'}\|_{\psi_1} \le 2 \sqrt{T} \sqrt{\sum_{i \in C_m} \sum_{j \in C_{m'}} s_{ij}^2} \|\epsilon\|_{\psi_2}^2.
\end{equation}
However, by assumption, $\sum_{i \in C_m} \sum_{j \in C_{m'}} s_{ij}^2 = 1$. Therefore we conclude
\begin{equation}
    \|\delta_{m,m'}\|_{\psi_1} \le 2 \sqrt{T} \|\epsilon\|_{\psi_2}^2
\end{equation}
and dividing by $\sqrt{T}$ on both sides yields the main result. The moment bound follows from Theorem 2.13 of \cite{Wainwright_2019}.
\end{proof}
\end{lemma}

We can also show that $\Delta$ is sub-exponential.

\begin{lemma}\label{lem::Delta_subexp} Under Assumptions \ref{assump::meanzero}-\ref{assump::subgaussian}, $\|\Delta\|_{\psi_1} \le 4 \|\epsilon\|_{\psi_2}^2 / \sigma_{\delta}$.
\begin{proof} By Eq. \eqref{eq::deltaM_comp}, we have
\begin{equation*}
    \Delta = \frac{1}{\sigma} \sum_{m, < m' \in [M]} \delta_{m, m'} = \frac{1}{\sigma} \sum_{i, j \in \mcI} s_{ij} \hat\epsilon_i^T (I_T - P) \hat\epsilon_j = \frac{1}{\sigma} \sum_{t, t' \in T} \sum_{m=1}^M \sum_{m'=1}^{m-1} \sum_{i \in C_m, j \in C_{m'}} s_{ij} ([I_T]_{t,t'} - P)_{t,t'} \hat\epsilon_{i,t} \hat\epsilon_{j,t'}.
\end{equation*}
This is a quadratic form in $\hat\epsilon$. By Proposition \ref{prop::subg2subexp} and Corollary \ref{cor::hateps_subG}, we conclude
\begin{align}
        \|\Delta\|_{\psi_1} 
    &\le 
        \frac{\|\epsilon\|_{\psi_2}^2}{\sigma} \sqrt{\sum_{t, t' \in T} \sum_{m=1}^M \sum_{m'=1}^{m-1} \sum_{i \in C_m, j \in C_{m'}} s_{ij}^2 ([I_T]_{t,t'} - P_{t,t'})^2} \\
    &=
        \frac{\|\epsilon\|_{\psi_2}^2}{\sigma} \sqrt{\sum_{m=1}^M \sum_{m'=1}^{m-1} \sum_{i \in C_m, j \in C_{m'}} s_{ij}^2 \|I_T - P\|_{\mathrm{Fr}}^2} \\
    &\le 
        \frac{\|\epsilon\|_{\psi_2}^2}{\sigma} \sqrt{4 \sum_{m'=1}^{m-1} \sum_{i \in C_m, j \in C_{m'}} s_{ij}^2} & \text{ since } \|I_T - P\|_{\mathrm{Fr}} \le 2 T \\
    &\le 
        \frac{2 \|\epsilon\|_{\psi_2}^2}{\sigma} \sqrt{\binom{M}{2} T} & \text{ since } \sum_{i \in C_m, j \in C_{m'}} s_{ij}^2 = 1.
\end{align}
However, Eq. \eqref{eq::sigma2_bound_useful} shows that $\sigma \ge 2 M \sqrt{T} \sigma_{\delta}$. Therefore we conclude
\begin{equation}
    \|\Delta\|_{\psi_1} \le \frac{4}{\sigma_{\delta}} \|\epsilon\|_{\psi_2}^2.
\end{equation}
\end{proof}
\end{lemma}

\subsection{Combinatorics results}

Fix $M, K \in \N$. Let $\mcM = \{(m, m') \in [M]^2 : m < m' \le m \}$. Fix $(g, h) \in \mcM^K$, so $g, h \in [M]^K$ and $g < h$ elementwise. Then we have the following definitions:
\begin{itemize}
    \item $\degree(g,h) \in [K]^{M}$ is the \textit{degree} vector satisfying $\degree_m(g,h) = \sum_{k=1}^K \I(g_k = m) + \I(h_k = m)$. 
    \item $\support(g,h) = \I(\degree(g,h) > 0) \in \{0,1\}^M$ denotes the \textit{support} vector which indicates elementwise whether cluster $m$ has positive degree.
    \item We say $(g,h)$ is a \textit{singleton} if there exists $m$ such that $\degree_m(g,h) =1$. We let $\singleton \defeq \{(g,h) : \exists m \text{ s.t. } \degree_m(g,h) = 1\}$ denote the set of singletons, and we let $\nonsingleton \defeq \mcM^K \setminus \singleton$ denote the set of non-singletons.
    \item We say that $(g,h)$ has \textit{even degree} if each coordinate of $\degree(g,h)$ is an even integer. We let $\mcE = \{(g,h) \in \mcM^K : \degree_m(g,h) \text{ is even for all } m \in [M]\}$ denote the set of even degree pairs $(g,h)$.
    \item Let $\mcE_m = \{(g,h) \in \mcE : \degree_m(g,h) > 0\}$ be the set of even $(g,h)$ such that cluster $m$ appears in $(g,h)$.
\end{itemize}

To prove our main results, we must bound $|\nonsingleton \setminus \mcE|$ and $|\mcE_m|$. To do this, we will split these sets into subsets based on the values of $\support(g,h)$ and $\degree(g,h)$. Thus, it is helpful to have the following notation.
\begin{enumerate}
    \item For any set $\mcA \subset \mcM^K$, $\bigsupport(\mcA) \defeq \{\support(g,h) : (g,h) \in \mcA\}$ denotes the set of support vectors compatible with $(g,h) \in \mcA$. Similarly, $\bigsupport_{\ell}(\mcA) = \{\support(g,h) : (g,h) \in \mcA, \sum_m \support_m(g,h) = \ell\}$ denotes the set of support vectors formed by $(g,h) \in \mcA$ such that there are exactly $\ell$ nonzero entries of $\support_m(g,h)$.
    \item  For any support vector $s \in \{0,1\}^M, \mcA \subset \mcM^K$, $\bigdegree(\mcA, s) \defeq \{\degree(g,h) : (g,h) \in \mcA, \support(g,h) = s\}$ denotes the set of degree vectors formed by $(g,h) \in \mcA$ satisfying $\support(g,h) = s$.
    \item For any degree vector $\delta \in [K]^M$, we let $\mcV(\mcA, \delta) = \{(g,h) \in \mcA : \degree(g,h) = \delta\}$ denote the elements of $\mcA$ compatible with this degree vector. 
\end{enumerate}
By splitting into cases, we can analyze the size of any subset $\mcA \subset \mcM^K$ by noting
\begin{equation}
    |\mcA| = \sum_{s \in \bigsupport(\mcA)} \sum_{\delta \in \bigdegree(\mcA, s)} |\mcV(\mcA, \delta)|
\end{equation}
or similarly
\begin{equation}\label{eq::split_by_degrees}
    |\mcA| = \sum_{\ell=1}^{M} \sum_{s \in \bigsupport_{\ell}(\mcA)} \sum_{\delta \in \bigdegree(\mcA, s)} |\mcV(\mcA, \delta)|.
\end{equation}
We now prove a pair of lemmas which are useful in analyzing the moments of our test statistic.

\begin{lemma}\label{lem::odd_nonsingleton_combo} There exists a universal constant $C(K)$ depending only on $K$ such that $|\nonsingleton \setminus \mcE| \le C(K) M^{K-1}.$
\begin{proof} We note that if $(g,h) \in \nonsingleton \setminus \mcE$, then $\degree(g,h)$ satisfies the following three properties:
\begin{itemize}
    \item $\sum_{m=1}^M \degree_m(g,h) = 2K$ since this is true of all $(g,h) \in \mcM^K$.
    \item $\degree(g,h) \ne 1$ holds elementwise by definition of $\nonsingleton$, i.e., $(g,h) \in \nonsingleton$ are not singletons.
    \item There exists some $m$ such that $\degree_m(g,h) \ge 3$, since $\degree(g,h)$ contains at least one odd number (since $(g,h) \not \in \mcE$) which is not equal to one.
\end{itemize}
Combining these three properties, we now claim that there can be at most $K-1$ nonzero entries of $\degree(g,h)$, i.e., $\|\degree(g,h)\|_0 \le K-1$. This is because all nonzero elements of $\degree(g,h)$ are greater than two and at least one equals three. Thus, $\sum_{m=1}^M \degree(g,h) \ge 2 \|\degree(g,h)\|_0 + 1$. But since $\sum_{m=1}^M \degree_m(g,h) = 2K$, we conclude $\|\degree(g,h)\|_0 \le \frac{2K-1}{2}$. Since $\|\degree(g,h)\|_0$ is an integer, this implies  $\|\degree(g,h)\|_0 \le K-1$. 

By definition, this means that $\bigsupport_{\ell}(\nonsingleton \setminus \mcE)$ is empty for all $\ell > K-1$. Therefore, if we set $\mcA \defeq \nonsingleton \setminus \mcE$, by Eq. (\ref{eq::split_by_degrees}) we learn
\begin{equation}
    |\mcA| \defeq |\nonsingleton \setminus \mcE| = \sum_{\ell=1}^{K-1} \sum_{s \in \bigsupport_{\ell}(\mcA)} \sum_{\delta \in \bigdegree(\mcA, s)} |\mcV(\mcA, \delta)|.
\end{equation}
Note that $\degree(g,h) = \delta$ fully determines the elements of $g,h$ up to their ordering. There are at most $(2K)!$ ways to order these elements among the $2K$ coordinates of $g,h$, and therefore $|\mcV(\mcA, \delta)| \le (2K)!$. Thus we note
\begin{equation}
    |\mcA| \le (2K)! \sum_{\ell=1}^{K-1} \sum_{s \in \bigsupport_{\ell}(\mcA)} |\bigdegree(\mcA, s)|.
\end{equation}
To measure $|\bigdegree(\mcA, s)|$, note that when $s$ has $\ell$ nonzero elements, this is bounded by the number of ways to distribute $2K$ indistinguishable objects (the degrees) into $\ell$ distinguishable bins (the nonzero coordinates of $s$), which is $\binom{2K+\ell-1}{2K}$. For $\ell \le K$, this is bounded by $\binom{3K}{2K}$. Thus we conclude
\begin{equation}
    |\mcA| \le (2K)! \binom{3K}{2K} \sum_{\ell=1}^{K-1} |\bigsupport_{\ell}(\mcA)|.
\end{equation}
Of course, $|\bigsupport_{\ell}(\mcA)|$ is bounded by the number of $M$-dimensional binary vectors with exactly $\ell$ coordinates equal to one, which is $\binom{M}{\ell}$. Therefore we conclude
\begin{equation}
    |\nonsingleton \setminus \mcE| \le (2K)! \binom{3K}{2K} \sum_{\ell=1}^{K-1} \binom{M}{\ell}.
\end{equation}
All terms are asymptotically negligible except the last in the sum. Therefore, there exists a universal constant $C(k)$ such that
\begin{equation}
    |\nonsingleton \setminus \mcE| \le C(k) M^{K-1}.
\end{equation}
\end{proof}
\end{lemma}

\begin{lemma}\label{lem::efronstein_combo} There exists a universal constant $C(K)$ depending only on $K$ such that $|\mcE_m| \le C(K) M^{K-1}$.
\begin{proof} For any $(g,h) \in \mcE_m$, $g$ and $h$ are both $K$-dimensional vectors, so $\sum_m \degree_m(g,h) = 2K$. Furthermore, $\degree_m(g,h)$ is always even, so this implies that $\support(g,h)$ has at most $K$ nonzero elements. Using Eq. \eqref{eq::split_by_degrees} and repeating a few arguments from the proof of Lemma \ref{lem::odd_nonsingleton_combo}, 
\begin{align*}
        |\mcE_m| 
    &\le 
        \sum_{\ell=1}^K \sum_{s \in \bigsupport_{\ell}(\mcE_m)} \sum_{\delta \in \bigdegree(\mcE_m, s)} |\mcV(\mcE_m, \delta)| \\
    &\le 
        (2K)! \sum_{\ell=1}^K \sum_{s \in \bigsupport_{\ell}(\mcE_m)} |\bigdegree(\mcE_m, s)| & \text{ since } |V(\mcE_m, \delta)| \le (2K)!\\
    &\le 
        (2K)! \binom{3K}{2K} \sum_{\ell=1}^K |\bigsupport_{\ell}(\mcE_m)| & \text{ since } |\bigdegree(\mcE_m, s)| \le \binom{3K}{2K} \text{ for } s \in \bigsupport_{\ell}(\mcE_m) \text{ with } \ell \le K.
\end{align*}
At this point, we note that $(g,h) \in \mcE_m$ implies that $\degree_m(g,h) > 0$ by definition. Therefore, if $\support(g,h)$ has $\ell$ nonzero elements, one of them must $m$, and there are at most $\binom{M-1}{\ell-1} \le \binom{M}{\ell-1}$ ways to choose the other nonzero coordinates. This implies that $|\bigsupport_{\ell}(\mcE_m)| \le \binom{M}{\ell-1}$ for each $\ell$, implying
\begin{equation*}
    |\mcE_m| \le (2K)! \binom{3K}{2K} \sum_{\ell=1}^{K} \binom{M}{\ell-1}.
\end{equation*}
All terms are asymptotically negligible as $M \to \infty$ except the last term. This implies that there exists some universal constant $C(K)$ such that
\begin{equation}
    |\mcE_m| \le C(K) M^{K-1}.
\end{equation}
\end{proof}
\end{lemma}

\subsection{Other miscallaneous lemmas}

\begin{lemma}\label{lem::uniform_ratio_consistency} Consider a triangular array of independent nonnegative random variables $X_1\upM, \dots, X_M\upM$ where $\mu_m\upM = \E[X_m\upM]$. Suppose there exists a constant $B > 0$ such that for some $\delta > 0$, (1) $\E[|X_m\upM - \mu_m\upM|^{1+\delta}] \le B$ and (2) $\mu_m\upM \ge B^{-1}$. Then
\begin{equation}
    \frac{\sum_{m=1}^M X_m\upM}{\sum_{m=1}^M \mu_m\upM} \toas 1.
\end{equation}
\begin{proof} A standard strong law of large numbers (for triangular arrays) shows that under condition (1), 
\begin{equation}
    \frac{1}{M} \sum_{m=1}^M X_m\upM - \frac{1}{M} \sum_{m=1}^M \mu_m\upM \toas 0.
\end{equation}
Since $\frac{1}{M} \sum_{m=1}^M \mu_m\upM \ge B^{-1}$ a.s. by condition (2), dividing by $\frac{1}{M} \sum_{m=1}^M \mu_m\upM$ yields that
\begin{equation}
    \frac{\frac{1}{M} \sum_{m=1}^M X_m\upM}{\frac{1}{M} \sum_{m=1}^M \mu_m\upM} - 1 \toas 0.
\end{equation}
This completes the proof.
\end{proof}
\end{lemma}

\end{document}